\documentclass[]{article}

\usepackage{arxiv}

\usepackage[utf8]{inputenc} 
\usepackage[T1]{fontenc}    
\usepackage{url}            
\usepackage{natbib}

\usepackage{hyperref}
\usepackage{booktabs}       
\usepackage{amsfonts}       
\usepackage{graphicx}

\usepackage{amsmath,amsthm,mathtools}
\usepackage{subcaption}
\usepackage{tikz}
\usetikzlibrary{arrows,positioning,calc}
\usepackage{tabularx,multirow}
\usepackage{enumitem}

\newtheorem*{example}{Example}

\title{Partially Specified Causal Simulations}

\date{12 September 2023}

\author{Alireza Zamanian\\
	Technical University of Munich,\\
	School of Computer, Information and Technology; \\
	Fraunhofer IKS, Germany\\
	\texttt{alireza.zamanian@iks.fraunhofer.de} \\
	\And
	Leopold Mareis\\
	Technical University of Munich,\\
	School of Computer, Information and Technology; \\
	Fraunhofer IKS, Germany\\
	\texttt{leopold.mareis@iks.fraunhofer.de} \\
	\And
	Narges Ahmidi \\
	Fraunhofer IKS, Germany\\
	\texttt{narges.ahmidi@iks.fraunhofer.de} \\
}

\begin{document}

\maketitle

\begin{abstract}
    Simulation studies play a key role in the validation of causal inference methods. The simulation results are reliable only if the study is designed according to the promised operational conditions of the method-in-test. Still, many causal inference literature tend to design over-restricted or misspecified studies. In this paper, we elaborate on the problem of improper simulation design for causal methods and compile a list of desiderata for an effective simulation framework. We then introduce partially randomized causal simulation (PARCS), a simulation framework that meets those desiderata. PARCS synthesizes data based on graphical causal models and a wide range of adjustable parameters. There is a legible mapping from usual causal assumptions to the parameters, thus, users can identify and specify the subset of related parameters and randomize the remaining ones to generate a range of complying data-generating processes for their causal method. The result is a more comprehensive and inclusive empirical investigation for causal claims. Using PARCS, we reproduce and extend the simulation studies of two well-known causal discovery and missing data analysis papers to emphasize the necessity of a proper simulation design. Our results show that those papers would have improved and extended the findings, had they used PARCS for simulation. The framework is implemented as a Python package, too. By discussing the comprehensiveness and transparency of PARCS, we encourage causal inference researchers to utilize it as a standard tool for future works.
\end{abstract}


\section{Introduction} \label{sec:introduction}

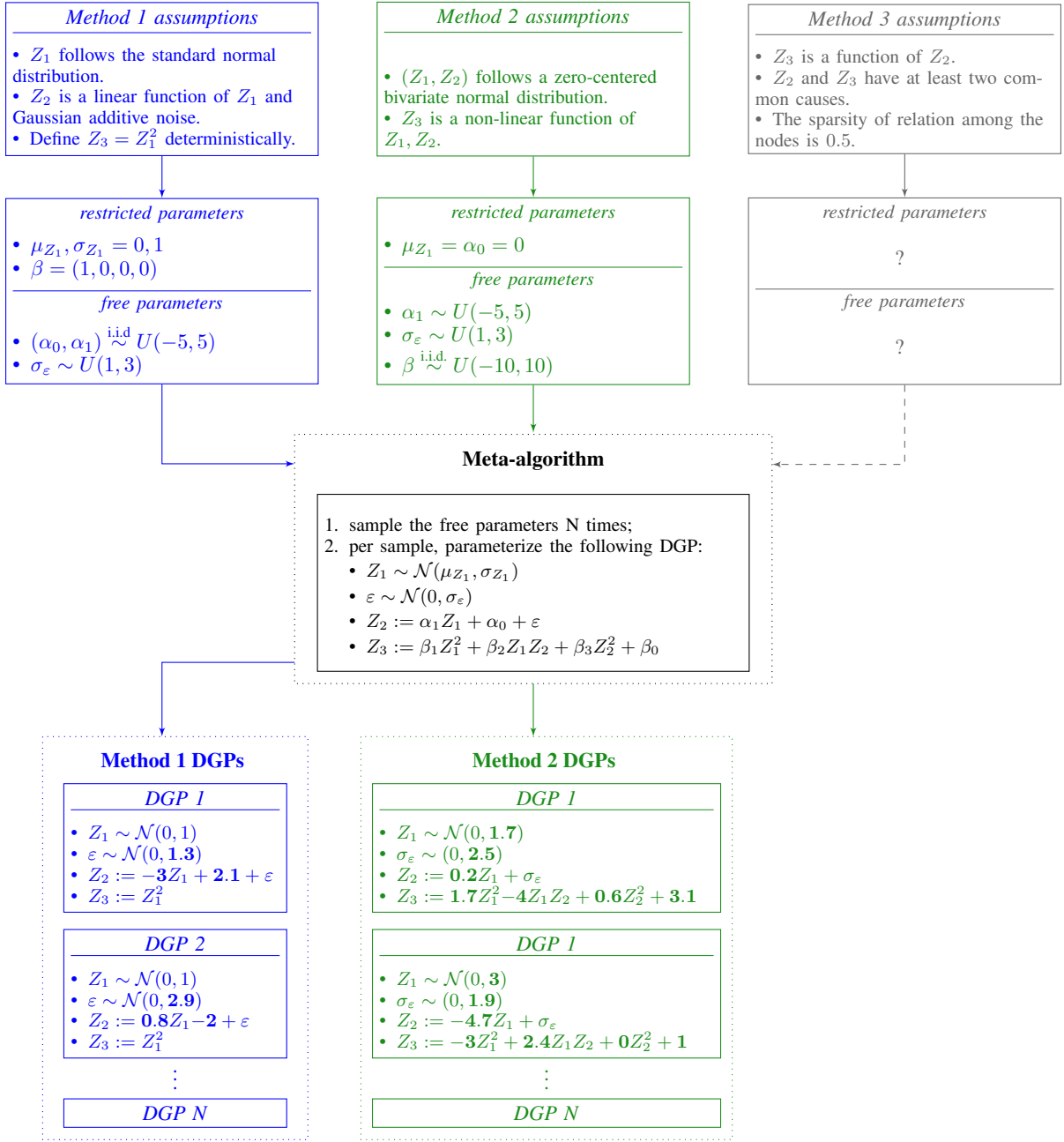
\begin{figure}[t]
    \centering
    \begin{tikzpicture}[scale=0.8, every node/.style={scale=0.9}]
        \tikzset{vertex/.style = {draw,circle}}
        \tikzset{edge/.style = {->,> = latex'}}
        \tikzset{block/.style = {draw}}
        
        \node[rectangle,
              draw,
              blue,
              text width=5cm,
              align=center] (paper1)  at (0, 0) {
              \textit{Method 1 assumptions} \\
              \vspace{0.2em}
              \hrule
              \vspace{0.2em}
              {\small \begin{itemize}[wide=0pt, noitemsep]
                  \item $Z_1$ follows the standard normal distribution.
                  \item $Z_2$ is a linear function of $Z_1$ and Gaussian additive noise.
                  \item Define $Z_3 = Z_1^2$ deterministically.
              \end{itemize}}};
        
        \node[rectangle,
              draw,
              green!50!black!90,
              text width=5cm,
              align=center] (paper2)  at (7, 0) {
              \textit{Method 2 assumptions} \\
              \vspace{0.2em}
              \hrule
              \vspace{1.2em}
              {\small \begin{itemize}[wide=0pt, noitemsep]
                  \item $(Z_1, Z_2)$ follows a zero-centered bivariate normal distribution.
                  \item $Z_3$ is a non-linear function of $Z_1, Z_2$.
              \end{itemize}}};
        
        \node[rectangle,
              draw,
              black!60,
              text width=5cm,
              align=center] (paper3)  at (14, 0) {
              \textit{Method 3 assumptions} \\
              \vspace{0.2em}
              \hrule
              \vspace{0.2em}
              {\small \begin{itemize}[wide=0pt, noitemsep]
                  \item $Z_3$ is a function of $Z_2$.
                  \item $Z_2$ and $Z_3$ have at least two common causes.
                  \item The sparsity of relation among the nodes is $0.5$.
              \end{itemize}}};

        \node[rectangle,
              draw,
              blue,
              text width=5cm,
              align=center] (params1) at (0, -4) {
              {\small \textit{restricted parameters}}\\
              \begin{itemize}[wide=0pt, noitemsep]
                  \item $\mu_{Z_1}, \sigma_{Z_1} = 0, 1$
                  \item $\beta = (1, 0, 0, 0)$
              \end{itemize}
              \hrule
              \vspace{0.2em}
              {\small \textit{free parameters}}\\
              \begin{itemize}[wide=0pt, noitemsep]
                  \item $(\alpha_0, \alpha_1) \stackrel{\text{i.i.d}}{\sim} U(-5, 5)$
                  \item $\sigma_\varepsilon \sim U(1, 3)$
              \end{itemize}};
        \draw[edge, blue] (paper1)  to (params1);
              
        \node[rectangle,
              draw,
              green!50!black!90,
              text width=5cm,
              align=center] (params2) at (7, -4) {
              {\small \textit{restricted parameters}}\\
              \begin{itemize}[wide=0pt, noitemsep]
                  \item $\mu_{Z_1} = \alpha_0 = 0$
              \end{itemize}
              \hrule
              \vspace{0.2em}
              {\small \textit{free parameters}}\\
              \begin{itemize}[wide=0pt, noitemsep]
                  \item $\alpha_1 \sim U(-5, 5)$
                  \item $\sigma_\varepsilon \sim U(1, 3)$
                  \item $\beta \stackrel{\text{i.i.d.}}{\sim} U(-10, 10)$
              \end{itemize}};
        \draw[edge, green!50!black!90] (paper2)  to (params2);
        
        \def\vsparam{\vspace{1.14em}}
        \node[rectangle,
              draw,
              black!60,
              text width=5cm,
              align=center] (params3) at (14, -4) {
              {\small \textit{restricted parameters}}\\
              \vsparam ? \vsparam
              \hrule
              \vspace{0.2em}
              {\small \textit{free parameters}}\\
              \vsparam ? \vsparam};
        \draw[edge, black!60] (paper3)  to (params3);

        \def\xmeta{7}
        \def\ymeta{-9.5}
        
        \draw[dotted] (\xmeta/2 -1, \ymeta + 2.8) rectangle ++(9, -4.7) {};
        \node[] (meta label) at (\xmeta, \ymeta + 2.3) {\textbf{Meta-algorithm}};
        
        \node[rectangle,
              draw,
              text width=7cm,
              align=center] (meta algorithm) at (\xmeta, \ymeta) {
              {\small
              \begin{enumerate}[wide=0pt, noitemsep]
                \item sample the free parameters N times;
                \item per sample, parameterize the following DGP:
                  \begin{itemize}
                      \item $Z_1 \sim \mathcal{N}(\mu_{Z_1}, \sigma_{Z_1})$
                      \item $\varepsilon \sim \mathcal{N}(0, \sigma_\varepsilon)$
                      \item $Z_2 := \alpha_1 Z_1 + \alpha_0 + \varepsilon$
                      \item $Z_3 := \beta_1 Z_1^2 + \beta_2 Z_1Z_2 + \beta_3 Z_2^2 + \beta_0$
                  \end{itemize}
              \end{enumerate}}
              \vspace{0.2em}};

        \def\xdgp{0}
        \def\ydgp{-12.5}
        \node[blue] (dgp label) at (\xdgp+0.2, \ydgp-0.35) {\textbf{Method 1 DGPs}};
        \draw[dotted, blue] (\xdgp-2.25, \ydgp+0.1) rectangle ++(5, -7.6) {};
        
        \node[rectangle,
            draw,
            text width=3.5cm,
            blue,
            align=center] (algo1)  at (\xdgp + 0.25, \ydgp-2) {
            \textit{DGP 1} \\
              \vspace{0.2em}
              \hrule
              \vspace{0.2em}
              {\small
              \begin{itemize}[wide=0pt, noitemsep]
                  \item $Z_1 \sim \mathcal{N}(0, 1)$
                  \item $\varepsilon \sim \mathcal{N}(0, \mathbf{1.3})$
                  \item $Z_2 := \mathbf{-3}Z_1 + \mathbf{2.1} + \varepsilon$
                  \item $Z_3 := Z_1^2$
              \end{itemize}}};
              
        \node[rectangle,
            draw,
            blue,
            text width=3.5cm,
            align=center] (algo2)  at (\xdgp + 0.25, \ydgp-4.75) {
            \textit{DGP 2} \\
              \vspace{0.2em}
              \hrule
              \vspace{0.2em}
              {\small
              \begin{itemize}[wide=0pt, noitemsep]
                  \item $Z_1 \sim \mathcal{N}(0, 1)$
                  \item $\varepsilon \sim \mathcal{N}(0, \mathbf{2.9})$
                  \item $Z_2 := \mathbf{0.8}Z_1 \mathbf{-2} + \varepsilon$
                  \item $Z_3 := Z_1^2$
              \end{itemize}}};
        \node[rectangle, blue,
            text width=3.5cm,
            align=center] (algo3)  at (\xdgp + 0.25, \ydgp - 6.25) {\vdots};
        \node[rectangle, blue,
            draw,
            text width=3.5cm,
            align=center] (algo4)  at (\xdgp + 0.25, \ydgp - 7) {\textit{DGP N}};
        
        \def\xdgp{7}
        \def\ydgp{-12.5}
        \node[green!50!black!90] (dgp label) at (\xdgp+0.2, \ydgp-0.35) {\textbf{Method 2 DGPs}};
        \draw[dotted, green!50!black!90] (\xdgp-3.25, \ydgp+0.1) rectangle ++(7, -7.6) {};
        
        \node[rectangle,
            draw,
            green!50!black!90,
            text width=5.6cm,
            align=center] (algo1)  at (\xdgp + 0.25, \ydgp-2) {
            \textit{DGP 1} \\
              \vspace{0.2em}
              \hrule
              \vspace{0.2em}
              {\small
              \begin{itemize}[wide=0pt, noitemsep]
                  \item $Z_1 \sim \mathcal{N}(0, \mathbf{1.7})$
                  \item $\sigma_\varepsilon \sim (0, \mathbf{2.5})$
                  \item $Z_2 := \mathbf{0.2}Z_1 + \sigma_\varepsilon$
                  \item $Z_3 := \mathbf{1.7}Z_1^2 \mathbf{-4} Z_1Z_2 + \mathbf{0.6}Z_2^2+ \mathbf{3.1}$
              \end{itemize}}};
              
        \node[rectangle,
            draw,
            green!50!black!90,
            text width=5.6cm,
            align=center] (algo1)  at (\xdgp + 0.25, \ydgp-4.75) {
            \textit{DGP 1} \\
              \vspace{0.2em}
              \hrule
              \vspace{0.2em}
              {\small
              \begin{itemize}[wide=0pt, noitemsep]
                  \item $Z_1 \sim \mathcal{N}(0, \mathbf{3})$
                  \item $\sigma_\varepsilon \sim (0, \mathbf{1.9})$
                  \item $Z_2 := \mathbf{-4.7}Z_1 + \sigma_\varepsilon$
                  \item $Z_3 := \mathbf{-3}Z_1^2 + \mathbf{2.4} Z_1Z_2 + \mathbf{0}Z_2^2+ \mathbf{1}$
              \end{itemize}}};
        \node[rectangle,
            green!50!black!90,
            text width=5.6cm,
            align=center] (algo3)  at (\xdgp + 0.25, \ydgp - 6.25) {\vdots};
        \node[rectangle,
            green!50!black!90,
            draw,
            text width=5.6cm,
            align=center] (algo4)  at (\xdgp + 0.25, \ydgp - 7) {\textit{DGP N}};

        \draw[edge, blue] (params1.south) -- ++(0, -1.5) -- ++(2.5, 0);
        \draw[edge, green!50!black!90] (params2.south) -- ++(0, -0.9);
        \draw[edge, dashed, black!60] (params3.south) -- ++(0, -1.5) -- ++(-2.5, 0);
        
        \draw[edge, blue]  (2.5, -11) -- ++(-2.5, 0) -- ++(0, -1.4);
        \draw[edge, green!50!black!90]  (7, -11.4) -- ++(0, -1);
            
    \end{tikzpicture}
    \caption{Schematic of 3 methods for validation (top row), a meta-algorithm (middle row) and generated DGPs for methods~1 and 2 (bottom row); The $Z$ variables are simulated study variables. Meta-parameters are: $\mu_{Z_1}, \sigma_{Z_1}, \sigma_\varepsilon, \alpha, \beta$. Restricted and free parameters are specified below each method. The bold face values in DGPs refer to the free parameters which are sampled. The meta-algorithm design does not suit the assumptions of method~3.}
    \label{fig:meta algo}
\end{figure}

Causal inference and discovery (CI), a crucial sub-field of statistics and artificial intelligence, aims to unravel cause-and-effect relationships from observational data. CI methods tackle a wide array of research questions of not associational, but causal nature, such as estimating the causal effect of a variable on another in a complex system, or discovering the underlying causal connections among study variables (See \cite{pearl2009causal}). These methods cannot be validated using observational data sets due to the unavailability of the ground truth. To mitigate this problem, simulation studies are widely used in the literature \citep{sofrygin2017simcausal, reisach2021beware}, where the ground truth is manually generated by the researchers and, therefore, accessible. At the core of simulation lies a data-generating process (DGP) comprising a set of equations and sampling procedures for variables, which determines the exact distribution of the synthesized data. Clearly, to generate reliable validation results, a DGP must be designed according to the promised operational conditions of the CI method, namely the mathematical assumptions. See \autoref{fig:meta algo} for examples of method assumptions and DGPs.

Generally, an assumption may lead to many DGPs that induce different data distributions. For instance, the ``non-linearity'' assumption for the variable relations (e.g., assumption over the variable $Z_3$ in \autoref{fig:meta algo}-Method~2) can take various forms, such as in second-order polynomial or trigonometric functions. For a complete validation, the method must be tested with all complying DGPs. The question is, how do we identify all complying DGPs? One solution is to devise a parametric algorithm (called ``meta-algorithm'' for brevity), with its internal parameters (``meta-parameters'') that outputs one DGP per each meta-parameter combination. If there is a clear mapping from the pool of assumptions to the meta-parameters (as is for \autoref{fig:meta algo}-Method~1), we identify and fix the subset of meta-parameters that relate to the assumptions (\textit{restricted parameters}). Then we sample the remaining ones (\textit{free parameters}) N times to generate N complying DGPs. This approach is adopted in the majority of CI literature for method validation; as notable attempts to design comprehensive meta-algorithms, we name the Simcausal R package \citep{sofrygin2017simcausal}, the simulation design for the 2017 Atlantic causal inference conference (ACIC) data analysis callenge \citep{hahn2019atlantic} and time series causal simulation framework by \cite{lawrence2021data}.

Designing a meta-algorithm for CI methods is a nontrivial and demanding task: firstly, meta-parameters must be relevant to a wide range of common assumptions in the field of application; otherwise, they cannot be adjusted accordingly. For example, in \autoref{fig:meta algo}, it is unclear how the meta-parameters can be adjusted to satisfy the assumptions of Method~3 about common causes. Secondly, the meta-parameters must be defined such that they can be fixed or varied independently; otherwise, the distinction between restricted and free parameters is difficult, if not impossible. For instance, the sparsity assumption in \autoref{fig:meta algo}-Method~3 requires the function coefficients such as $\alpha$ and $\beta$ to be adjusted only jointly. Lastly, the design must be comprehensive and inclusive so that the meta-algorithm itself does not cause a harmful restriction on the validation results. For example, the meta-algorithm in \autoref{fig:meta algo} is improperly restricted to only three variables, and the non-linearity of the relations is only defined for one variable ($Z_3$) and only via the second-order polynomial form.

Scanning the literature, we noticed that simulation studies often suffer from improperly designed meta-algorithms. This causes the generated DGPs to be considerably limited to few specific scenarios and do not cover the entire operational space of the method. Consequently, the validation results become incomplete, unreliable, and, in extreme cases, misleading.

In this paper, we gather a list of common causal assumptions found in literature and compile a set of desiderata for a potentially effective simulation framework. We then introduce PARCS (\textbf{pa}rtially \textbf{r}andomized \textbf{c}ausal \textbf{s}imulation), a simulation meta-algorithm aimed to satisfy those desiderata and address the aforementioned limitations. We explain how PARCS relates to the common causal assumptions, controls disentangled and independent simulation parameters, and provides comprehensive complying DGPs for various collections of assumptions. We believe PARCS allows researchers to overcome the issues of simulation design, run more in-depth analyses of their methods, and report their simulation setup more transparently and reproducibly. PARCS is implemented and available as a Python package\footnote{The Python package is available at \url{https://pypi.org/project/pyparcs/}. A detailed documentation is available at \url{https://fraunhoferiks.github.io/parcs/}} for researchers.

The main contributions of this paper are listed below.
\begin{enumerate}
    \item We present the causal methods simulation desiderata based on the common simulation studies found in the CI literature.

    \item We introduce PARCS simulation meta-algorithm. We show how common causal assumptions map to PARCS parameters. We present the governing equations, design choices, and theoretical details for the simulation process.
    
    \item We demonstrate the utility of PARCS by reproducing and extending the simulation studies done by two SoA CI papers by \citet{jarrett2022hyperimpute} and \citet{shimizu2006linear}. Throughout experiments, we show how PARCS generates additional CI validation scenarios which were untested and ignored by the papers, and thus allows for more extensive and precise validation.
\end{enumerate}

The rest of the paper is organized as follows: A review of common causal assumptions in literature is presented in \autoref{sec:desiderata}. \autoref{sec:theoretical_background} provides the theoretical background and explains the design of PARCS. Simulation case studies can be found in \autoref{sec:case study}. We revisit the planned simulation desiderata and present the conclusion in \autoref{sec:conclusion}.

\section{Simulation desiderata} \label{sec:desiderata}

Let the random vector $Z \in \mathbb{R}^d$ represent the study variables in a data set. The joint density of $(Z_1, \dots, Z_d)$ can be factorized as a product of one-dimensional conditional densities, as
\begin{align} \label{eq:dag factor}
    P(z) = \prod\nolimits_i P(z_{\gamma(i)} \;| \;s^\gamma(i)), \quad
    s^\gamma(i) = \{z_{\gamma(1)}, \dots, z_{\gamma(i-1)}\}
\end{align}
for any permutation $\gamma$ over the components $\{1, \dots, d\}$ where $S^\gamma(i)$ is the set of all variables preceding $i$ in the $\gamma$ ordering. Furthermore, we define intervention on a component $Z_j \in Z$ as altering the conditional density $P(Z_j|.)$ in \autoref{eq:dag factor} to $Q(Z_j|.)$, where $Q$ represents the probability distributions after intervention. In a basic case, $Q(Z_j|.)$ represents a degenerate distribution; this is equal to setting $Z_j$ to a fixed value $Z_j=z^*_j$. It is clear that after intervention, the joint density over $Z$ (and perhaps some of the one-dimensional conditional densities) changes, i.e., $P(z) \neq Q(z)$. 

\autoref{eq:dag factor} offers a simulation procedure, by which we specify and sequentially sample from the conditional densities along the order of $\gamma$. Although this procedure is applicable to any permutation, we are interested in permutations with a specific causal property; in particular, a permutation $\gamma^c$, under which interventions on $Z_{\gamma^c(j)}$ does not change the distributions of the upstream nodes $S^{\gamma^c}(j)$. As a result, the corresponding distribution for an intervention defined by $Q$ is given as
\begin{align} \label{eq:intrv}
   Q(z) = 
   \prod_{i=1}^{j-1} P(z_{\gamma^c(i)}|s^{\gamma^c}(i)) \cdot
   \prod_{i=j}^{d} Q(z_{\gamma^c(i)}|s^{\gamma^c}(i)) \cdot
\end{align}
In terms of simulation procedure, \autoref{eq:intrv} reads as ``\textit{sample for the upstream variables from the pre-intervention, and for the intervened and the downstream variables from the post-intervention densities}''.

This leads us to the definition of a causal directed acyclic graph (DAG): a DAG $\mathcal{G}$ is defined over a node set ${N = \{Z_1, \dots, Z_d \}}$ corresponding to the study variables, and a directed edge set ${E \subseteq \{ (i_1, i_2) \in N^2: i_1 \in  S^{\gamma^c}(i_2)\}}$. Each edge represents a causal effect from one node (parent) to another (child). By definition, only the parent nodes affect the child node. Denoting the set of all parent nodes for $Z_i$ in a DAG by $\mathrm{pa}_\mathcal{G}(i)$, we rewrite \autoref{eq:dag factor} as $P(z) = \prod_{i=1}^d P(z_i \;| \; \mathrm{pa}_\mathcal{G}(i))$. This means that given the parent set, a node is independent of the rest of the graph\footnote{known as the Markov property of DAGs}.

By the definition of causal DAG, the causal ordering $\gamma^c$ is the topological sorting in DAG. This implies ${\mathrm{pa}_\mathcal{G}(i) \subseteq S^{\gamma^c}(i)}$, namely, the parents of a node precede it in the sorting. To sample from a DAG, based on the general sampling procedure for \autoref{eq:dag factor}, we sample the nodes along its topological sorting, and use the realizations at each step as the inputs of the next step. An illustrative example of a simple DAG is depicted in \autoref{fig:simple causal dag}. In this DAG, $Z_1$ is a parent of $Z_2$, and $Z_1, Z_2$ are parents of $Z_3$. An arbitrary permutation can be $\gamma = \{1, 3, 2\}$ which leads to the factorization $P(Z_1)P(Z_3|Z_1)P(Z_2|Z_1, Z_3)$. On the other hand, given the causal permutation $\gamma^c = \{1, 2, 3\}$, the distribution factorizes as $P(Z_1)P(Z_2|Z_1)P(Z_3|Z_1, Z_2)$. Assuming an intervention on $Z_3$, the last two terms under $\gamma$, and only the last term under $\gamma^c$ changes. 

In order to fully specify the statistical model for a DAG, it is sufficient to assume a working model for each conditional density. These models are defined in form of a structural causal model (SCM) as
\begin{align} \label{eq:sem}
    Z_i = h_i(Z_{\mathrm{pa}_\mathcal{G}(i)}, \varepsilon_i), \quad i \in \{1, \dots, d \},
\end{align}
via a deterministic functions $h_i(.): \mathbb{R}^{|\mathrm{pa}_\mathcal{G}(i)| + 1} \to \mathbb{R}$ and independent random errors $\varepsilon_i$. To sample from an observational distribution as in \autoref{eq:sem}, we sample error terms and calculate the variables' realization along the topological order $\gamma$ of the DAG.

Equations~\ref{eq:dag factor} and \ref{eq:sem} introduce the necessary and sufficient set of parameters that define a causal model of DAG and SCM: the graph's nodes $Z$ and edges $E$, deterministic functions $h(.)$ and the distributions of the noise terms $\varepsilon$. Causal methods make their assumptions on one or more of these parameters, partially or fully, directly or indirectly. \autoref{tab:ci literature} outlines a (non-exhaustive) list of assumptions which are central to the CI literature, and their relationship to the four introduced parameters. Based on \autoref{tab:ci literature}, we present the following list of desiderata which must be statisfied by a potential CI simulation framework in order to be effective and useful for the CI community.

\begin{table}
    \centering
    \begin{tabularx}{\textwidth}{c p{0.3\textwidth} p{0.55\textwidth}}
        \hline
        \textit{Parameter} & \textit{Assumptions} & \textit{Example literature} \\
        \toprule
        $Z$ & number of nodes${}^{*}$ & \\
            & expected value, hidden/observed &\cite{shpitser2013counterfactual, mohan2013missing, mohan2022graphical}\\ [0.5cm]
        $E$ & DAG structure & \cite{spirtes1991algorithm, uhler2013geometry, shpitser2013counterfactual, pfister2022identifiability, xu2021learning, mohan2013missing, mohan2022graphical} \\
            & sparsity & \cite{spirtes1991algorithm, uhler2013geometry, peters2015structural, pfister2022identifiability} \\
            & linearity of relations${}^{**}$ & \cite{spirtes1991algorithm, shimizu2006linear, hoyer2008nonlinear}, \\ [0.5cm]
        $h$ & governing distribution & \cite{hoyer2008nonlinear} \\ 
            & model coefficients & \cite{pfister2022identifiability} \\ [0.5cm]
        $\varepsilon$ & noise distribution and model & \cite{hoyer2008nonlinear, peters2015structural} \\
                      & noise independence${}^{***}$ & \\
        \bottomrule
    \end{tabularx}
    \vspace{1em}
    \caption{Examples of DAG-related assumptions in CI literature. The lists of sub-categories and papers are non-exhaustive. \\
    *  The number of nodes is often not a direct assumption of the methods, but almost all the simulation studies wrongly fix the number of nodes in DGP. \\
    ** The mentioned papers explicitly put an assumption on linearity/non-linearity of the relations. \\
    *** Almost all CI methods make an implicit noise-independence assumption.}
    \label{tab:ci literature}
\end{table}

\paragraph{A. Underlying causality} As the first and most fundamental requirement, the framework must be able to embody the notion of causality and govern the simulation based on causal relations. To satisfy this requirement, it is sufficient to take the causal DAG and SCM models (Equation~\ref{eq:dag factor} and~\ref{eq:sem}) as the DGP.

\paragraph{B. Data dimensions and modality} The framework must provide means to control the statistical characteristics of data such as sample size, number of variables, data types and conditional densities. Furthermore, selecting the distribution of any variable must not be limited by other simulation parameters.

\paragraph{C. Controlled complexity} Data complexity must be controlled in terms of number of variables, causal relations, non-linearity of relations, higher-order polynomial terms, and causal parents interactions of the causes of a variable. Different sources of complexity should be independently controllable, such that the framework allows us to induce complexity by adjusting each source as desired.

\paragraph{D. Noise models} The framework must provide means to induce various types of noise in the data. The noise profile must be free to be affected and controlled by graph variables.

\section{PARCS simulation meta-algorithm} \label{sec:theoretical_background}

To design a meta-algorithm that satisfies Desideratum A, we assume the underlying DGP to be a causal DAG and an accompanying SCM, and adopt the sampling procedures defined by Equations~\ref{eq:dag factor}-\ref{eq:sem}. In addition, we employ a set of design choices for $h$ and $\varepsilon$ to satisfy Desiderata B, C, and D. These choices lead to disentanglement of the parameters and thus allow for restricted and free parameters distinction. We introduce this concept in the current section denoted by \textit{partial randomization}.

\subsection{DAG sampling procedure} \label{sec:sampling from causal graph}

Take an arbitrary node $Z_i$ and let the $j$ index selects the node indices in $\mathrm{pa}(i)$. Given the samples of $Z_{\mathrm{pa}(i)}$, the sampling procedure for $Z_i$ is as follows.

\paragraph{Edge functions} Samples of each $Z_j$ are transformed by a so-called \textit{edge function} $e_{ji}: \mathbb{R} \to \mathbb{R}$ as
\begin{align} \label{eq:edge function}
    Z_{j \rightarrow i} = e_{ji}(Z_j; \phi_{ji}), \quad \text{ for }  j \in \mathrm{pa}(i),
\end{align}
where $\phi_{ji}$ parameterizes $e_{ji}$. Edge functions may introduce the first source of non-linearity in the simulation process. Being reminiscent of the generalized linear model's \textit{link function} concept, each type of functions (such as log, logistic or Gaussian radial basis functions) induces a particular pattern in the parent-child relation. Following the notation in \autoref{eq:edge function}, we denote the vector of the transformed samples of all parents (by their corresponding edge functions) for $Z_i$ as $Z_{\mathrm{pa}(i) \rightarrow i}$.

\paragraph{Input library} We assume a library $\zeta(x)$ consisting of any desired form of linear and non-linear combinations of elements of a vector $x$. For instance, $\zeta: x \mapsto (x_1^2, x_2^2, \dots, x_d^2), \; x \in \mathbb{R}^d$ is a library that returns the 2\textsuperscript{nd}-order polynomials. We apply this library to the edge-transformed samples of the parents, i.e. $Z_{\mathrm{pa}(i) \rightarrow i}$, to obtain an input vector for $Z_i$ (hence the name, \textit{input library}). As we will show in the next step, this vector is used to specify the distribution of $Z_i$.

The complexity of the input library $\zeta$ is predicated on the variety of the functions in $\zeta$, which may induce trigonometric or n\textsuperscript{th}-order polynomial terms. These transformations represent the additional source of non-linearity in the simulation process. Including each form must be justified by the designer of the simulation study. As a basic choice, we consider the following input library for the rest of the paper\footnote{\autoref{eq:zeta} is the choice also in the PARCS Python library, at least up to version 1.0.0. It may change in future versions.}:
%

%
\begin{subequations}
\begin{align} \label{eq:zeta}
    \zeta \; :& \; x \mapsto \left(1, \zeta_\text{linear}\left(x\right), \zeta_\text{quad.}\left(x\right)\right), \quad x \in \mathbb{R}^d,
\end{align}
where
\begin{align}
    \zeta_\text{linear} \; :& \; x \mapsto  x, \\ \nonumber
    \zeta_\text{quad.} \; :& \; x \mapsto (x_i x_j : x_i, x_j \in x).
\end{align}
\end{subequations}

\paragraph{Parameterized distributions} By the probability integral transform theorem, a random variable $X$ follows a distribution $P$ with the inverse cumulative distribution function (iCDF) $F_P^{-1}$, if $X = F_P^{-1}(\varepsilon)$ and an error $\varepsilon$ follows a continuous uniform distribution $\text{Unif}(0, 1)$. We employ this theorem for choosing the structural equation function $h$ for the nodes. More specifically, we choose $h_i(.)$ to be the iCDF of a probability distribution $P^i_\theta$, parameterized by $\theta$. The parent nodes affect $Z_i$ via controlling the distribution parameters: for the parameter space $\Theta^i$, a coefficient matrix $W^i \in \mathbb{R}^{ |\Theta^i| \times |\zeta^i|}$ is specified and the distribution parameters are given by
\begin{align} \label{eq:param linear}
    \theta^i = W^i \zeta(Z_{\mathrm{pa}(i) \to i})^{\top}.
\end{align}
By sampling the error term as $\varepsilon_i \sim \text{Unif}(0, 1)$ and calculating the $Z_i$ realizations as $ Z_i = h_i(\varepsilon_i; \theta^i) $, we have
\begin{align}
    Z_i \; | \; Z_{\mathrm{pa}(i)} \sim P^i_\theta,
\end{align}
namely, conditioned on its parents, a node follows the distribution whose iCDF is selected for $h_i$. The choice of iCDF results in different statistical characteristics, data types and noise models in the synthesized data. The values of $W^i$ also determine which of the distribution parameters (and how) the parents affect in the child node. For instance, all-zero coefficients for one of the distribution parameters means that the parents do not affect the distribution via that parameter.

In summary, a DGP in PARCS is specified by sets of output distributions and parameter coefficient vectors of its nodes, and transforming functions of its edges. The sampling procedure repeats for all nodes in the topological (causal) order of the graph. For source nodes (nodes with no parents), the corresponding coefficient vector $W$ is a vector of length $|\theta^i|\times 1$, including the bias terms (considering $\zeta$ in \autoref{eq:zeta}). \autoref{tab:meta-algorithm parameters} summarizes the parameters of the PARCS simulation meta-algorithm.

\begin{table}[h]
    \centering
    \begin{tabularx}{0.9\textwidth}{c c l l}
        \hline
        \textit{Parameter} & \textit{Example value} & \textit{Description} & \textit{Controls} \\
        \toprule
        $|Z|$ & 5 & number of nodes & DAG structure \\
        $E$ & $\{(Z_1,Z_2), \dots\}$ & edges & DAG structure, sparsity\\
        $P^i_\theta$ & Bern, Unif, $\mathcal{N}$ , \dots & node distributions & data type, noise model, expectations \\
        $\zeta$ & $(1, Z^2_{j\rightarrow i}, \dots)$ & input library & non-linearity \\
        $W^i$ & $(1, 0.2, \dots)$ & coefficient vectors for $\theta$ & causal relations  \\
        $e_{ji}$ & $\arctan(Z; \alpha, \beta, \gamma)$ & edge functions for edges & non-linearity \\
        $\phi_{ji}$ & $\alpha=1.5, \dots$ & edge function parameters & non-linearity \\
        \hline
    \end{tabularx}
    \vspace{1em}
    \caption{PARCS meta-parameters. The last 5 parameters must be specified for all nodes/edges.}
    \label{tab:meta-algorithm parameters}
\end{table}

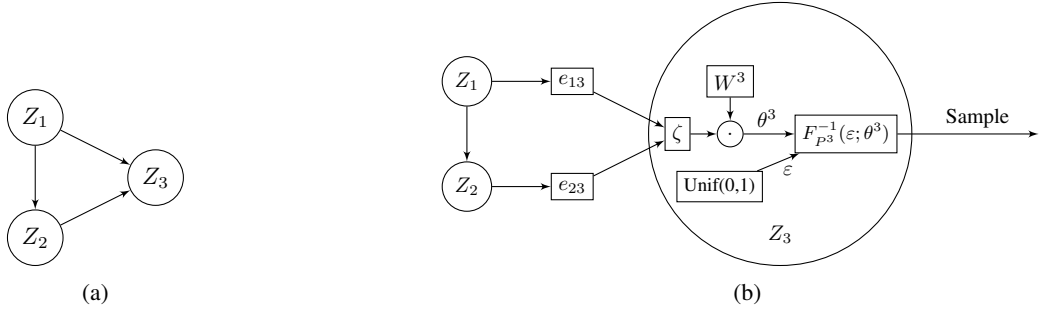
\begin{figure}[t]
    \centering
    \begin{subfigure}[b]{0.4\textwidth}
        \centering
        \begin{tikzpicture}[scale=0.8, every node/.style={scale=0.9}]
            \tikzset{vertex/.style = {draw,circle}}
            \tikzset{edge/.style = {->,> = latex'}}
            \tikzset{block/.style = {draw}}
            
            \node[vertex] (z1) at (-3,0) {$Z_1$};
            \node[vertex] (z2) at (-3,-2) {$Z_2$};
            \node[vertex] (z3) at (-1,-1) {$Z_3$};
            
            \draw[edge] (z1) to (z3);
            \draw[edge] (z2) to (z3);
            \draw[edge] (z1) to (z2);
        \end{tikzpicture}
        \caption{}
        \label{fig:simple causal dag}
    \end{subfigure}
    \hfill
    \begin{subfigure}[b]{0.55\textwidth}
    \centering
    \begin{tikzpicture}[scale=0.7, every node/.style={scale=0.8}]
        \tikzset{vertex/.style = {draw,circle}}
        \tikzset{edge/.style = {->,> = latex'}}
        \tikzset{block/.style = {draw}}
        
        \node[vertex] (z1) at (0,1) {$Z_1$};
        \node[vertex] (z2) at (0,-1) {$Z_2$};
        
        \node[block] (e13) at (2,1) {$e_{13}$};
        \node[block] (e23) at (2,-1) {$e_{23}$};
        
        \node[block] (zeta) at (4,0) {$\zeta$};
        \node[block] (W) at (5,1) {$W^3$};
        \node[vertex] (dotprod) at (5,0) {$.$};
        
        \node[block] (N) at (7.2,0) {\small $F^{-1}_{P^3}(\varepsilon;\theta^3)$};
        \node[block] (error) at (4.8,-1) {\small Unif(0,1)};
        \node[] (out) at (11, 0) {};
        
        \draw[] (5.95, 0) circle (2.5);
        \node[] at (5.95, -1.9) {$Z_3$};

        \draw[edge] (z1) to (e13);
        \draw[edge] (z2) to (e23);
        \draw[edge] (z1) to (z2);
        
        \draw[edge] (e13) to (zeta);
        \draw[edge] (e23) to (zeta);
        
        \draw[edge] (zeta) to (dotprod);
        \draw[edge] (W) to (dotprod);
        
        \draw[edge] (dotprod) to (N);
        \draw[edge] (error) to (N);
        \draw[edge] (N) to (out);
        \node[] at (9.7, 0.3) {Sample};
        \node[] at (5.7, 0.3) {$\theta^3$};
        \node[] at (6.1, -0.7) {$\varepsilon$};
        
    \end{tikzpicture}
    \caption{}
    \label{fig:parcs mechanics}
    \end{subfigure}
    \caption{Example of sampling procedure for a simple DAG with three nodes $Z_1, Z_2, Z_3$: (a) the corresponding causal DAG, and (b) PARCS mechanics to sample $Z_3$ values.}
\end{figure}

\begin{example}
    
\autoref{fig:simple causal dag} illustrates a simple DAG with three nodes ($Z_1, Z_2, Z_3$): a normally distributed node $Z_3$ and its two parents, $Z_1, Z_2$ (following arbitrary distributions). Assume all edge functions are identity functions. For $Z_3$, we have $Z_{\mathrm{pa}(3)} = (Z_1, Z_2)$. We transform the parents' samples as
\( Z_{1\rightarrow 3} = I(Z_1) = Z_1, \; Z_{2\rightarrow 3} = I(Z_2) = Z_2 \).
The input vector (following \autoref{eq:zeta}) is $\zeta(Z_{\{1, 2\}\rightarrow3}) = \left(1, Z_1, Z_2, Z_1^2, Z_1Z_2, Z_2^2\right)^\top$. We choose $P^3$ to be the Gaussian normal $\mathcal{N}(\mu, \sigma)$. We have $\Theta^3 = \mathbb{R} \times \mathbb{R}_{>0}, \theta^3 = (\mu, \sigma) \in \Theta^3 $.  Let the coefficient matrix be
\(
        W^3 = \left( \begin{smallmatrix} W^3_\mu \\ W^3_\sigma \end{smallmatrix} \right) = \left(\begin{smallmatrix} 1 & 1 & 1 & 0 & 0 & 0 \\ 2 & 0 & 0 & 0 & 1 & 0 \end{smallmatrix} \right),
\)
which gives the following for $Z_3$ in the DGP:
\[
    Z_3|z_1, z_2 \sim \mathcal{N}(\mu=z_1+z_2+1, \;\; \sigma=z_1z_2+2).
\]
For example $ Z_3|Z_1=0.2, Z_2=0.3 \sim \mathcal{N}(\mu = 1.5, \sigma = 2.06)$. Note that sampling from $\mathcal{N}$ is done by sampling the error terms uniformly and passing it to $F^{-1}_{\mathcal{N}}$, parameterised by $\theta^3$. The described process is depicted in \autoref{fig:parcs mechanics}.
\end{example}

The sampling procedure extends to interventional distributions by a simple adjustment according to \autoref{eq:intrv}. Intervention on a causal model is defined changing the conditional distribution of a node away from its natural (observational) conditional distribution. Intervention can be done by setting the realizations of a node to a constant value (degenerate distribution), or changing the SCM parameters, output distribution or the parent set of the node. Regardless of the type of intervention, we repeat the same sampling procedure as above, along the causal ordering of the DAG, until we reach a node (or nodes) $Z_i$ on which we intervene. Then, the normal sampling is interrupted, and the realizations are set according to the intervention. Afterwards, the normal procedure again takes over; however, note that the intervened values now are passed to the descendants\footnote{PARCS Python package allows for in-place intervention actions for the DGP. With the help of reusable error terms, counterfactual analysis is possible.}.

\subsection{Partial randomization} \label{sec:partial randomization}

As described in \autoref{sec:introduction} and depicted by \autoref{fig:meta algo}, generating DGPs based on method assumptions is done by identifying the restricted meta-parameters, and varying the free meta-parameters. We argued why independence among meta-parameters is crucial in this process. In this section, we explain how this process, called partial randomization for brevity, is implemented in the PARCS meta-algorithm.

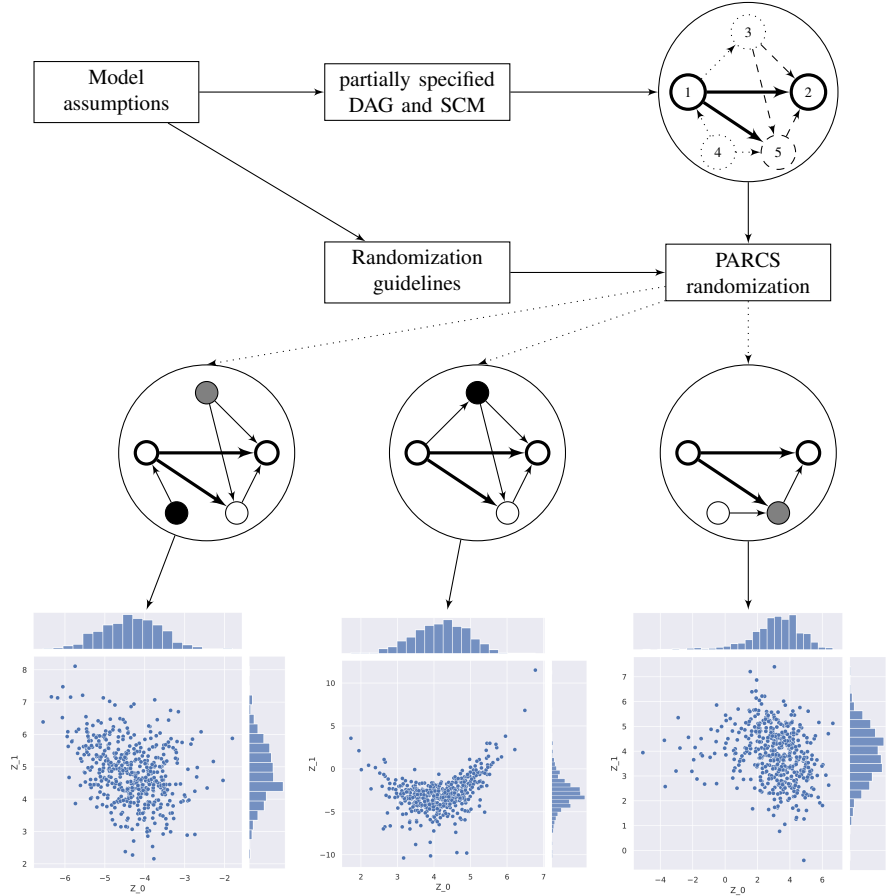
\begin{figure}[t]
    \centering
    \begin{tikzpicture}[scale=0.8, every node/.style={scale=0.9}]
        \tikzset{vertex/.style = {draw,circle}}
        \tikzset{edge/.style = {->,> = latex'}}
        \tikzset{block/.style = {draw}}
        
        \node[rectangle,
              draw,
              text width=2.2cm,
              align=center] (assump) at (-5,0) {\small Model assumptions};
        \node[rectangle,
              draw,
              text width=2.5cm,
              align=center] (desc) at (0,0) {\small partially specified DAG and SCM};
        \node[draw, circle, thin, minimum size=2.65cm] (graph) at (5.5, 0) {};
        \draw[edge] (assump) to (desc);
        \draw[edge] (desc) to (graph);

        \node[vertex, very thick] (n1) at (4.5, 0) {\tiny 1};
        \node[vertex, very thick] (n2) at (6.5, 0) {\tiny 2};
        \node[vertex, dotted] (n3) at (5.5, 1) {\tiny 3};
        \node[vertex, dotted] (n4) at (5, -1) {\tiny 4};
        \node[vertex, dashed] (n5) at (6, -1) {\tiny 5};
        
        \draw[edge, very thick] (n1) to (n2);
        \draw[edge, very thick] (n1) to (n5);
        
        \draw[edge, dotted] (n1) to (n3);
        \draw[edge, dashed] (n3) to (n2);
        \draw[edge, dotted] (n4) to (n1);
        \draw[edge, dotted] (n4) to (n5);
        \draw[edge, dashed] (n3) to (n5);
        \draw[edge, dashed] (n5) to (n2);

        \node[rectangle,
              draw,
              text width=2.5cm,
              align=center] (guideline) at (0,-3) {\small Randomization guidelines};
        \node[rectangle,
              draw,
              text width=2.2cm,
              align=center] (parcs) at (5.5,-3) {\small PARCS randomization};
        \draw[edge] (graph) to (parcs);
        \draw[edge] (guideline) to (parcs);
        \draw[edge] (assump) to (guideline);
        
        
        \def\x{5.5}
        \def\y{-6}
        \node[vertex, very thick] (n1) at (\x-1, \y) {};
        \node[vertex, very thick] (n2) at (\x+1, \y) {};
        \node[vertex] (n4) at (\x-0.5, \y-1) {};
        \node[vertex, fill=gray] (n5) at (\x+0.5, \y-1) {};
        \node[draw, circle, thin, minimum size=2.6cm] (graph1) at (\x, \y) {};
        
        \draw[edge, very thick] (n1) to (n2);
        \draw[edge, very thick] (n1) to (n5);
        
        \draw[edge] (n4) to (n5);
        \draw[edge] (n5) to (n2);
        
        
        \def\x{1}
        \def\y{-6}
        \node[vertex, very thick] (n1) at (\x-1, \y) {};
        \node[vertex, very thick] (n2) at (\x+1, \y) {};
        \node[vertex, fill=black] (n3) at (\x, \y+1) {};
        \node[vertex] (n5) at (\x+0.5, \y-1) {};
        \node[draw, circle, thin, minimum size=2.6cm] (graph2) at (\x, \y) {};
        
        \draw[edge, very thick] (n1) to (n2);
        \draw[edge, very thick] (n1) to (n5);
        
        \draw[edge] (n1) to (n3);
        \draw[edge] (n3) to (n2);
        \draw[edge] (n3) to (n5);
        \draw[edge] (n5) to (n2);
        
        
        \def\x{-3.5}
        \def\y{-6}
        \node[vertex, very thick] (n1) at (\x-1, \y) {};
        \node[vertex, very thick] (n2) at (\x+1, \y) {};
        \node[vertex, fill=gray] (n3) at (\x, \y+1) {};
        \node[vertex, fill=black] (n4) at (\x-0.5, \y-1) {};
        \node[vertex] (n5) at (\x+0.5, \y-1) {};
        \node[draw, circle, thin, minimum size=2.6cm] (graph3) at (\x, \y) {};
        
        \draw[edge, very thick] (n1) to (n2);
        \draw[edge, very thick] (n1) to (n5);
        
        \draw[edge] (n3) to (n2);
        \draw[edge] (n4) to (n1);
        \draw[edge] (n3) to (n5);
        \draw[edge] (n5) to (n2);
        
        \draw[edge, dotted] (parcs) to (graph1.north);
        \draw[edge, dotted] (parcs) to (graph2.north);
        \draw[edge, dotted] (parcs) to (graph3.north);
        
        \node[] (dist1) at (5.5, -11) {\pgftext{\includegraphics[width=120pt]{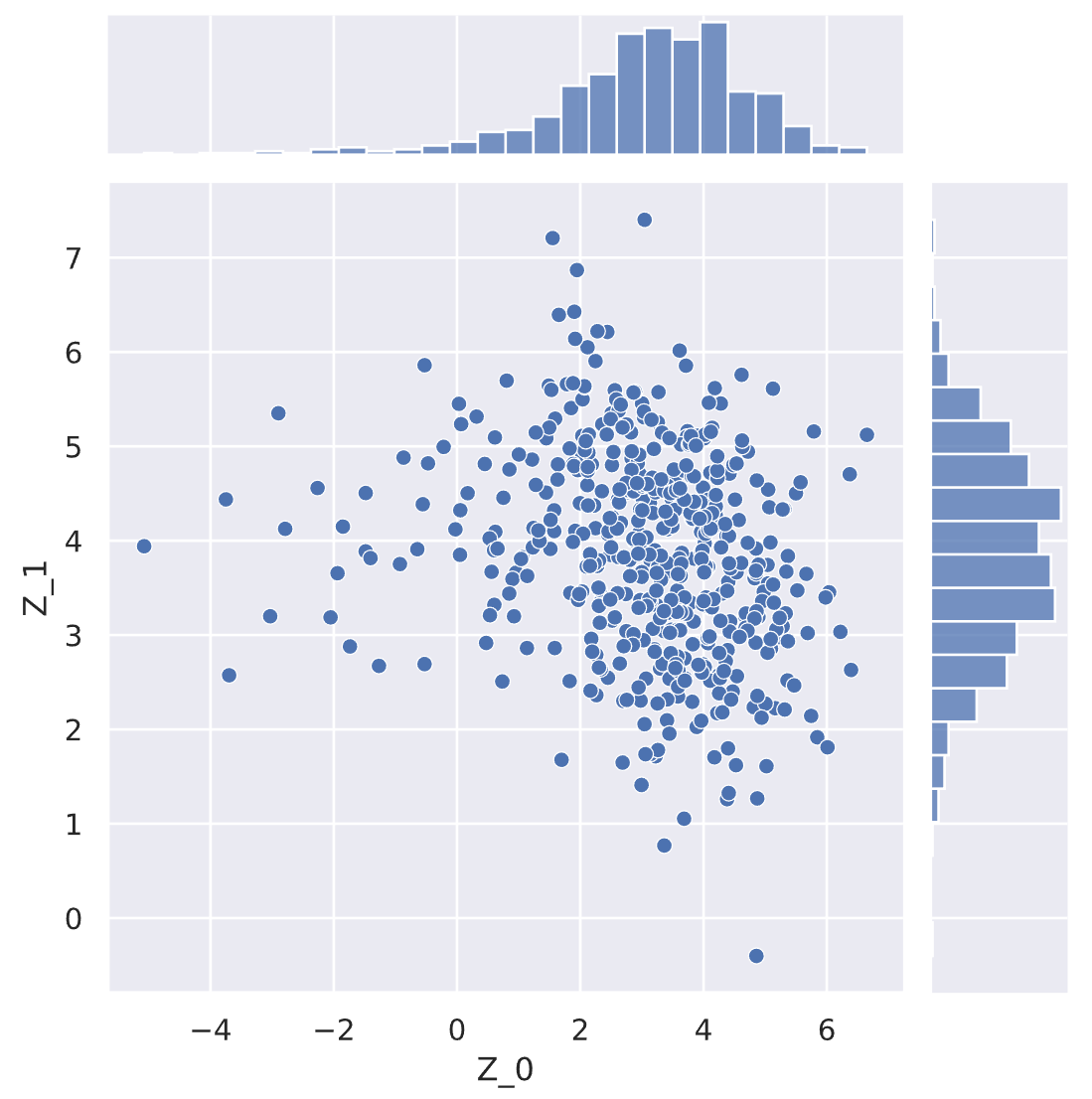}}};
        \node[] (dist2) at (0.5, -11) {\pgftext{\includegraphics[width=120pt]{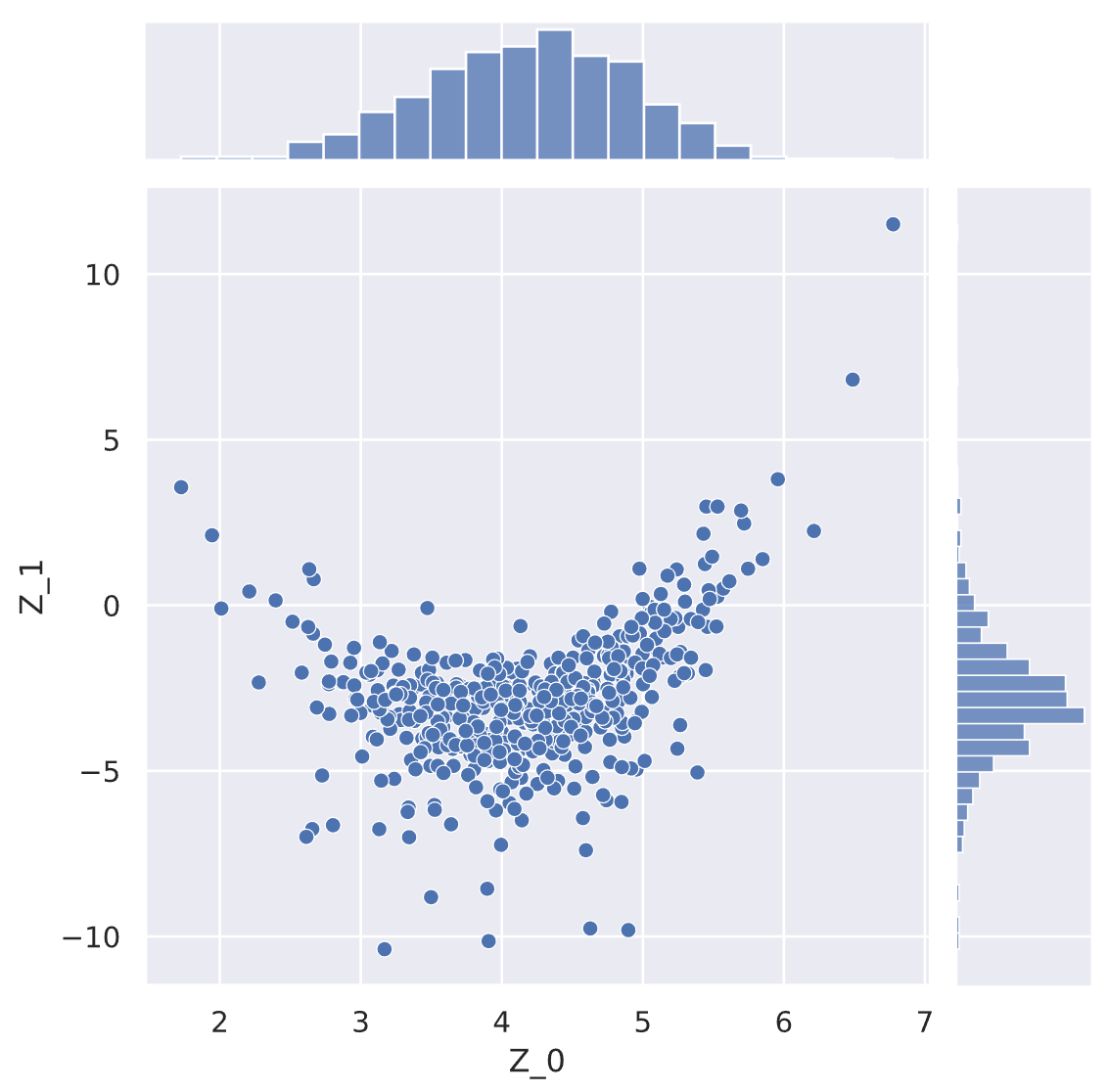}}};
        \node[] (dist3) at (-4.5, -11) {\pgftext{\includegraphics[width=120pt]{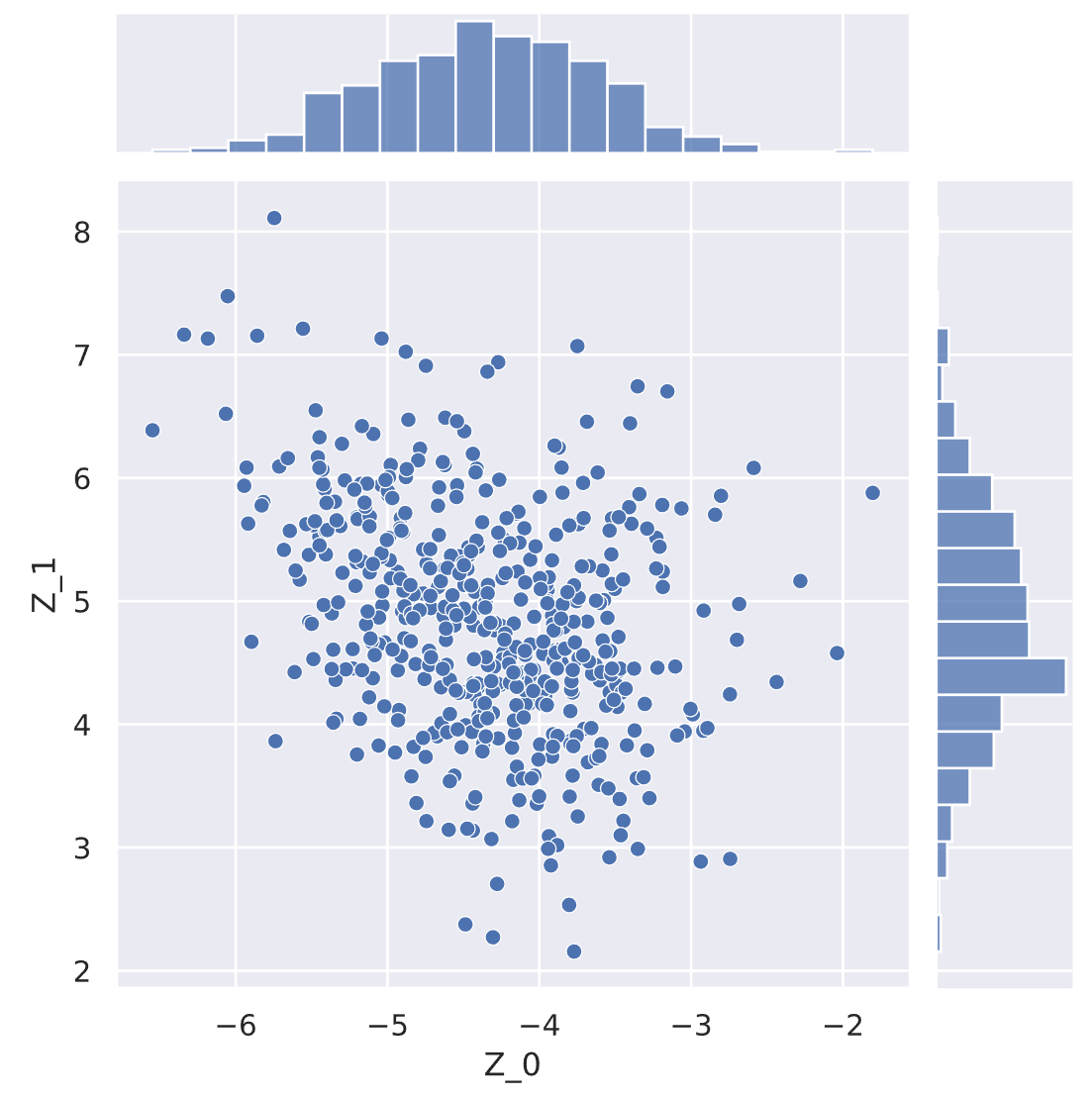}}};
        
        \draw[edge] (graph1) to ($(dist1.north)+(0, 2.25)$);
        \draw[edge] (graph2) to ($(dist2.north)+(0, 2.25)$);
        \draw[edge] (graph3) to ($(dist3.north)+(0, 2.25)$);
        \node[] (end) at (0, -13.5) {};
        
    \end{tikzpicture}
    \caption{A schematic of graph randomization in PARCS. thick nodes and edges are fully specified by the user. Dashed and dotted nodes and edges are partially specified and randomized. In the resulting three graphs, distributions and edge functions may differ (represented by different colors). All graphs comply with the specified sub-graph, and vary in the undefined sub-parts.}
    \label{fig:partial randomization}
\end{figure}

The meta-parameters in PARCS follow a natural order, from the broadest to the narrowest scope, which starts by the overall DAG structure and ends with all the edge functions and node distributions along with their parameters. However, users do not need to follow this order when restricting the meta-parameters. they can leave any of the meta-parameters up to randomization, and restrict the rest, regardless of the rank of the parameter in the order. \autoref{fig:partial randomization} depicts a schematic of the partial randomization process via for a hypothetical scenario. The partially specified DGP (top right) represents the model assumptions. In this DAG, the existence of three nodes (1, 2, 5) is assumed. While node 1 and 2 have a fixed distribution, node 5 awaits the randomization process for its distribution (hence depicted by dashed line). Furthermore, assumptions imply that there may exist a node 3 (hence depicted by dotted line) which might receive an edge from 1 (dotted $1\rightarrow3$ edge), but if it exists, then it definitely have outgoing edges to 2 and 5 (dashed $3\rightarrow2$ and $3\rightarrow 5$ edges), namely it can be a confounder for 2 and 5. Finally, a node 4 may exist, and it may be the parent of node 1, node 5, both or none (dotted $4\rightarrow1$ and $4\rightarrow5$ edges).

According to the meta-parameter order, we first randomly decide the existence of nodes 3 and 4. If the assumptions concern the chance of having these nodes, they must be reflected in the randomization guidelines. For instance, the guidelines might impose a success probability of $0.5$ for node 3, and $0.8$ for node 4. After each randomization, we have a new DGP where the existence of nodes 3 and 4 has been decided. The three sampled DAGs in the \autoref{fig:partial randomization} (bottom row) present different results for these nodes after randomization.

Next, we place the new edges based on the status of the newly added nodes and the randomization guidelines. Edges $3\rightarrow5$ and $3\rightarrow2$ must be added if node $3$ exists (as mandated by the assumptions), while we must randomize the existence of $1\rightarrow3$, $4\rightarrow1$ and $4\rightarrow5$. This can be done, for instance, by randomly deciding on the existence of edges according to a sparsity parameter. Sparsity can be either determined by the model assumptions, or vary freely in a given range. Subsequently, the edge functions and their parameters are randomized (or fixed by the user).

Finally, we randomize the distribution for node $5$ and the newly-added nodes. As before, we choose the distributions ($P$) from a given list of options in the randomization guideline, and then sample the coefficients ($W$) based on the number of parents and the input library ($\zeta$) that is specified by the guideline. Note that this step of randomization is not limited by the meta-parameters of the previous steps by any means. Even though the number of sampled coefficients (trivially) depend on the number of parent nodes, type of the distribution, complexity of the input library, and ranges of $W$ can be freely randomized. Furthermore, every coefficient can be specified directly by the user, not to be left for randomization. For example, the user can fix the causal coefficient of node $1$ for the mean parameter of $2$ beforehand, and randomize the rest of the coefficients.

While the example in \autoref{fig:partial randomization} might seem disconnected to the actual examples in the CI literature, it showcases several scenarios that may arise in real use cases. As hinted in \autoref{tab:ci literature}, the actual CI assumptions translate to the four DAG/SCM parameters whatsoever, and those are being handled by PARCS meta-parameters. To clarify the partial randomization process further, we revisit the method assumptions in \autoref{fig:meta algo} and explain briefly, how PARCS generates DGPs for each:

\begin{example}

Simulation studies for the three method assumptions in \autoref{fig:meta algo} can be implemented by PARCS as follows ($\zeta$ is selected based on \autoref{eq:zeta}):

\begin{itemize}
    \item[\textbf{M1}] \begin{enumerate}
        \item Set $Z = \{Z_1, Z_2, Z_3\}, E = \{(Z_1,Z_2), (Z_1, Z_3) \} $;
        \item set $e_{12}, e_{13}$ to identity function;
        \item set $P^1$ to normal; $W^1_\mu = 0$, $W^1_\sigma = 1$;
        \item set $P^2$ to normal; randomize question marks in $W^2_\mu = (?, ?, 0)$ (suppressing the $Z_1^2$ term); randomize $W^2_\sigma = (?, 0, 0)$ (std of the error independent of $Z_1$);
        \item set $P^3$ to normal; set $W^3_\mu = (0, 0, 1)$ and $W^3_\sigma = (0, 0, 0)$.
    \end{enumerate}
    
    \item[\textbf{M2}] \begin{enumerate}
        \item Set $Z = \{Z_1, Z_2, Z_3\}, E = \{(Z_1,Z_2), (Z_1, Z_3), (Z_2, Z_3) \} $;
        \item set $Z_1$, $Z_2$ nodes and $(Z_1, Z_2)$ edge parameters as in M1;
        \item randomize $e_{13}$ and $e_{23}$ by selecting from available edge functions; randomize $\phi_{13}$ and $\phi_{23}$;
        \item set $P^3$ to normal with zero std; randomize $W^3_\mu$ fully.
    \end{enumerate}
    
    \item[\textbf{M3}] \begin{enumerate}
        \item Randomize the number of common causes for $Z_2, Z_3$ and create the new nodes.
        \item Randomize an adjacency matrix for the new nodes according to the desired sparsity level.
        \item Set and randomize $Z_2$, $Z_3$, newly sampled nodes and all edges, similar to the process explained for M1 and M2.
    \end{enumerate} 
\end{itemize}

\end{example}

\subsection{Refined simulations}

A challenge in the partial randomization process is to produce valid and meaningful DAG configurations. In particular, the randomized DGPs shall not violate the valid range of the distribution parameters, must lead to the desired variability in the samples, and shall not produce simulation artifacts which open a door to spurious inference on the synthesized data. Below, we explain two possible options in PARCS design that users can choose to rely on, in order to address these challenges.

\subsubsection{Parameter restrictions and controlled variability}

Many distribution parameters, such as Bernoulli's success probability, are defined in bounded ranges. Since, the partial randomization process samples the free parameters independently and ignores the statistical characteristics of the resulting structural equations, it is conceivable how a combination of parameters might lead to invalid ranges for the distribution parameters. For example, suppose ${Z_2 \sim \text{Bern}(p=2Z_1)}$. Any DGP that does not impose ${Z_1 \in [0, 0.5]}$, will result in invalid success probability values for $Z_2$. The issue can be resolved by imposing extra restrictions on the SCM, e.g. forcing $Z_1$ to follow a uniform distribution with ranges smaller than $[0, 0.5]$; however, this defies the very purpose of PARCS, which is limiting the DGP only according to the model assumptions.
\begin{subequations}

As a solution, we propose the use of post-processing \textit{correction} functions. Applying to nodes (distributions), a correction operator transforms the values of the distribution parameters to a valid range. This is done via a Sigmoid function, applied to the raw values of the distribution parameters as
\begin{align} \label{eq:node correction}
    f_\text{corr.}(\theta; L, U) = \frac{U-L}{1+e^{-\theta}} + L,
\end{align}
where $L$ and $U$ are the desired lower and upper bounds of the transformed values. \autoref{eq:node correction} guarantees that for all the sampled meta-parameters, the distribution parameters receive the inputs in their respective valid ranges. Note that by applying correction, the parent-child nodes relations are inevitably non-linear. Nevertheless, such transformations are common in statistical modeling, e.g. in the logistic models, where the binary variable is defined as $y = \sigma(w^\top x)$.

While \autoref{eq:node correction} guarantees the validity of the sampled parameters, the quality of resulting data is not controlled. Following the argument of correction function, consider the child node with Bernoulli distribution ${Z_2 \sim \text{Bern}(p=2Z_1)}$ and the parent node defined as $Z_1 \sim \mathcal{N}(\mu=10, \sigma=1)$. After correction, the success probability is spread around $0.9995$ with the standard deviation of $1.54\times10^{-5}$ (empirical estimation over $1000$ samples). This means that $Z_2$ is heavily imbalanced toward $1$. While this scenario could be desired and intended in some designs, in general, we want to simulate data with reasonable (controlled) variability. This can be achieved by introducing an offset parameter in the correction function, and revising \autoref{eq:node correction} as 
\begin{align} \label{eq:node correction target mean}
    f_\text{corr.}(\theta; L, U, O) = \frac{U-L}{1+e^{-\theta + O}} + L.
\end{align}
By controlling the offset, we shift the inputs to the correction function, and hence change the mean of the output. In PARCS, rather than choosing $O$ explicitly, we set a \textit{target mean} parameter. This way, the corresponding offset is estimated empirically over a selected number of burn-in samples upon instantiating the graph; it will remain fixed for the next sampling rounds.

\end{subequations}

\subsubsection{Mitigating the simulation artifacts}

Distribution parameters are directly affected by the statistics of the parent nodes. In particular, the measurement unit and variance of the parent determines the characteristics of the child. This is an undesired simulation effect which leads to invalidity of the simulation process. \cite{reisach2021beware} shows how data scale and marginal variance might undesirably carry information about the simulation DAG structure. Moreover, in an extreme case where the support of a parent is considerably larger than the others, it will dominate the matrix product in \autoref{eq:param linear}. As a result, \textit{smaller magnitude} parents lose  their effect solely due to unit of measurement. To address the issue, we introduce a post-processing \textit{edge correction} function which is a simple standardization of the edge inputs by applying
\begin{align} \label{formula_correction}
    f_{\text{edge corr.}}(Z_j; \mu, \sigma) = \frac{Z_j - \mu}{\sigma},
\end{align}
where $\mu$ and $\sigma$ are empirical mean and standard deviation of $Z_j$. Similar to the offset argument in node correction function, the moments are calculated empirically over the burn-in samples upon instantiating the graph and will remain fixed for the next sampling rounds. By enabling correction, sampled causal distributions are prevented from establishing invalid structural equations and produce more meaningful simulations.

\subsubsection{More complex simulations}

No meta-algorithm design covers all possible scenarios; the goal must be to design the capabilities and limitations according to the field of research. PARCS is no exception to this rule. While it is designed to be effective for the assumptions in \autoref{tab:ci literature}, there are certain limitations in the design. Particularly, the polynomial form of the distribution parameters, i.e. Equations~\ref{eq:zeta} and \ref{eq:param linear}, cannot contain functions of multiple parents, e.g., ${\theta_k = \sin{(Z_1+Z_2)} + Z_3}$ (the form could have been implemented via the choice of the edge function if the arbitrary function is applied to one parent). This form, however, is still possible to induce for PARCS, using an intermediate `dummy' variable. In this case, we can imagine a new (deterministic) variable is defined as $X = Z_1 + Z_2$, while the $e_{XZ}$ edge function is the sine edge function. The message is, despite the apparent incapability to simulate complex DGPs, PARCS provides indirect methods which we can leverage in certain situations, while retaining the same randomization options.

\subsection{Comparison with other frameworks}

The majority of CI research papers which run simulation studies for empirical validations, employ a hard-coded and limited DGP, similar to the example in \autoref{fig:meta algo}. In addition, benchmarking data sets have been introduced for causal methods, such as the pairwise causal discovery data sets \citep{guyon2019cause}. Regardless of the complexity and diversity of scenario, both options lack the flexibility for extension beyond a pre-defined DGP. Thus, for a meaningful comparison of our method with existing solutions, we draw redears' attentions to two simulation frameworks which, similar to PARCS, are designed to be applicable in various CI scenarios, but have limited funcitonalities.

Simcausal \citep{sofrygin2017simcausal} is a programming package, designed to facilitate defining and sampling procedures of SCMs. Simcausal can prove effective in designing causal simulations, not only in static, but also in temporal settings. The drawback of the simcausal is that it requires full specification of the structural equations for all variables. This means that researchers must take extra steps and implement a parameterization and randomization procedure in order to simulate partially specified DAGs and SCMs. PARCS overcomes this limitation via the partial randomization functionality, which is possible because of its meta-parameter design. In summary, simcausal's meta-algorithm does not go beyond the basic SCM simulation and sampling procedures.

The next framework, and the closest to the PARCS design, is the causal simulation framework for time series, proposed by \cite{lawrence2021data} (which we call `temporal framework' for ease of reference).  In this framework, time series simulation is based on sequential DAGs, in which the state of each temporal variable at each time point is represented by a distinct node. The framework's meta-parameters are mainly tailored for a unique temporal structure with `feature', `target', and `latent' variables. For all nodes, users can define (or randomize) the number of parents or children, complexity of the distributions, and noise variances. Furthermore, the temporal aspect of the DAG is controllable via parameters such as the `maximum temporal lag' in the parent-child pairs. Temporal framework also introduces refinement procedures to prevent exploding of variables after recursive matrix multiplications.

Similar to PARCS, users of the temporal framework can partially specify a sequential DAG, and randomize the undefined meta-parameters. We, however, argue that PARCS' meta-parameters and randomization procedure proves more flexible and adjustable to various method assumptions. Firstly, the temporal framework cannot be adjusted to DAG structures beyond the feature-target-latent structure, e.g. the instrumental variable setting or missing graphs (see \autoref{sec:missingness}). For many structures, the framework does not provide relevant meta-parameters, while users are forced to specify or randomize a number of existing parameters which are irrelevant, e.g., the lag parameter for static DAGs).

Secondly, the simulation complexity is adjusted via only one meta-parameter, while, as argued in the desideratum C, there are many sources of complexity such as non-linear equations, non-linear edge functions or non-additive non-Gaussian dependent noise terms. In summary, the temporal framework can prove effective for a range of simulation studies, especially for time series methods, but will fall short in a wide range of studies which can be handled by PARCS\footnote{Beyond the meta-algorithm design, the programming package for PARCS and the temporal framework can be compared in terms of utility, user-friendliness, transparency, and documentation. We leave this comparison to future works, as it is out of the scope of this paper. It is worthy of note, that the PARCS package provides temporal causal simulation based on the same `sequential DAG' idea.}.

\section{Case studies} \label{sec:case study}

In this section, we reproduce and extend the simulation studies of two SoA papers to showcase the advantages of PARCS and its power to enhance the CI studies. First paper by \citet{jarrett2022hyperimpute} introduces Hyperimpute, an missing data imputation method for ignorable missingness mechanisms in complex datasets. Next, we look in the field of causal inference at the effective and widely-used statistical LiNGAM method \citep{shimizu2006linear}. This method recovers the causal order and the causal linear coefficients. Beside the popularity and wide usage, we selected these two works because the validity of their methods depends significantly on their assumptions, i.e. DAG structure and non-linearity assumptions for Hyperimpute and linearity assumption for LiNGAM.

\subsection{Hyperimpute: a missing data analysis case study} 
\label{sec:missingness}

\begin{figure}[t]
    \centering
    \begin{subfigure}[b]{0.25\textwidth}
        \centering
        \begin{tikzpicture}[scale=0.5, every node/.style={scale=0.7}]
            \tikzset{vertex/.style = {draw,circle}}
            \tikzset{edge/.style = {->,> = latex'}}
            \tikzset{block/.style = {draw}}
            
            \node[vertex] (z1) at (0,0) {$Z_1$};
            \node[vertex] (z2) at (2,0) {$Z_2$};
            \node[vertex] (z3) at (4,0) {$Z_3$};
            
            \draw[edge] (z1) to (z2);
            \draw[edge] (z2) to (z3);
            \draw[edge] (z1) to[bend left] (z3);
            
            \node[vertex] (r1) at (0,-2) {$R_1$};
            \node[vertex] (r2) at (2,-2) {$R_2$};
            \node[vertex] (r3) at (4,-2) {$R_3$};
        \end{tikzpicture}
        \caption{MCAR mechanism}
        \label{fig:m-graph mcar}
    \end{subfigure}
    \begin{subfigure}[b]{0.25\textwidth}
        \centering
        \begin{tikzpicture}[scale=0.5, every node/.style={scale=0.7}]
            \tikzset{vertex/.style = {draw,circle}}
            \tikzset{edge/.style = {->,> = latex'}}
            \tikzset{block/.style = {draw}}
            
            \node[vertex] (z1) at (0,0) {$Z_1$};
            \node[vertex] (z2) at (2,0) {$Z_2$};
            \node[vertex] (z3) at (4,0) {$Z_3$};
            
            \draw[edge] (z1) to (z2);
            \draw[edge] (z2) to (z3);
            \draw[edge] (z1) to[bend left] (z3);
            
            \node[vertex] (r2) at (2,-2) {$R_2$};
            \node[vertex] (r3) at (4,-2) {$R_3$};
            
            \draw[edge] (z1) to (r2);
            \draw[edge] (z1) to (r3);
        \end{tikzpicture}
        \caption{MAR mechanism}
        \label{fig:m-graph mar}
    \end{subfigure}
    \begin{subfigure}[b]{0.25\textwidth}
        \centering
        \begin{tikzpicture}[scale=0.5, every node/.style={scale=0.7}]
            \tikzset{vertex/.style = {draw,circle}}
            \tikzset{edge/.style = {->,> = latex'}}
            \tikzset{block/.style = {draw}}
            
            \node[vertex] (z1) at (0,0) {$Z_1$};
            \node[vertex] (z2) at (2,0) {$Z_2$};
            \node[vertex] (z3) at (4,0) {$Z_3$};
            
            \draw[edge] (z1) to (z2);
            \draw[edge] (z2) to (z3);
            \draw[edge] (z1) to[bend left] (z3);
            
            \node[vertex] (r1) at (0,-2) {$R_1$};
            \node[vertex] (r2) at (2,-2) {$R_2$};
            \node[vertex] (r3) at (4,-2) {$R_3$};
            
            \draw[edge] (z1) to (r2);
            \draw[edge] (z1) to (r3);
        \end{tikzpicture}
        \caption{MNAR mechanism}
        \label{fig:m-graph mnar}
    \end{subfigure}
    
    \begin{subfigure}[b]{0.4\textwidth}
        \centering
        \begin{tikzpicture}[scale=0.5, every node/.style={scale=0.7}]
            \tikzset{vertex/.style = {draw,circle}}
            \tikzset{edge/.style = {->,> = latex'}}
            \tikzset{block/.style = {draw}}
            
            \node[vertex] (z1) at (0,0) {$Z_1$};
            \node[vertex] (z2) at (2,0) {$Z_2$};
            \node[vertex] (z3) at (4,0) {$Z_3$};
            
            \draw[edge] (z1) to (z2);
            \draw[edge] (z2) to (z3);
            \draw[edge] (z1) to[bend left] (z3);
            
            \node[vertex] (r1) at (0,-2) {$R_1$};
            \node[vertex] (r2) at (2,-2) {$R_2$};
            \node[vertex] (r3) at (4,-2) {$R_3$};
            
            \draw[edge] (z1) to (r1);
            \draw[edge] (z2) to (r2);
            \draw[edge] (z3) to (r3);
        \end{tikzpicture}
        \caption{self-censoring (MNAR) mechanism}
        \label{fig:m-graph sc}
    \end{subfigure}
    \begin{subfigure}[b]{0.4\textwidth}
        \centering
        \begin{tikzpicture}[scale=0.5, every node/.style={scale=0.7}]
            \tikzset{vertex/.style = {draw,circle}}
            \tikzset{edge/.style = {->,> = latex'}}
            \tikzset{block/.style = {draw}}
            
            \node[vertex] (z1) at (0,0) {$Z_1$};
            \node[vertex] (z2) at (2,0) {$Z_2$};
            \node[vertex] (z3) at (4,0) {$Z_3$};
            
            \draw[edge] (z1) to (z2);
            \draw[edge] (z2) to (z3);
            \draw[edge] (z1) to[bend left] (z3);
            
            \node[vertex] (r1) at (0,-2) {$R_1$};
            \node[vertex] (r2) at (2,-2) {$R_2$};
            \node[vertex] (r3) at (4,-2) {$R_3$};
            
            \draw[edge] (z1) to (r2);
            \draw[edge] (z1) to (r3);
            
            \draw[edge] (z2) to (r1);
            \draw[edge] (z2) to (r3);
            
            \draw[edge] (z3) to (r1);
            \draw[edge] (z3) to (r2);
        \end{tikzpicture}
        \caption{no-self-censoring (MNAR) mechanism*}
        \label{fig:m-graph nsc}
    \end{subfigure}
    \caption{Five examples of m-graphs, modeling different missingness mechanisms: MCAR (a), MAR (b), MNAR (c, d, e). *The term \textit{no-self-censoring} is coined by \citet{malinsky2022semiparametric} for a similar graph as in (e), but with undirected edges among $R$ nodes, hence modeling a chain graph. In theory, the graph in (e) is considered a submodel.}
    
    \label{fig:m-graph examples}
\end{figure}
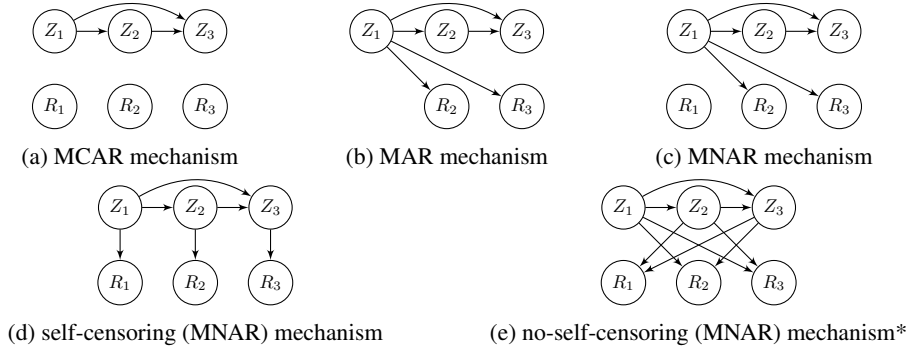

Missing data can be posed as a causal inference problem via \textit{m-graph} models~\citep{mohan2013missing, shpitser2015missing, mohan2022graphical}. An m-graph is a causal DAG with $Z$ and $R$ nodes, representing the main variables and missingness indicators respectively. The $Z \rightarrow R$ relations determine the missingness mechanism. \autoref{fig:m-graph examples} presents five examples of m-graphs, inducing various missing data mechanisms. Being a causal DAG, m-graphs can be simulated using PARCS. To showcase this ability, we analyze the simulation study in \citet{jarrett2022hyperimpute} for the Hyperimpute imputation algorithm. We briefly describe the paper's simulation setting, then continue with reproducing and extending some of the presented results using PARCS. We demonstrate that authors have correctly reported the results of a subset of scenarios which complies with their assumptions, but ignore a wide range of complying scenarios still. We show that the proposed method achieves considerably different performances in some of the untested scenarios, which means the analysis in the paper is incomplete due to simulation limitations. For ease of reference, we refer to this paper as `Hyperimpute paper'.

\subsubsection{Original simulation setting} 
\label{sec:sim setting}
In the Hyperimpute paper simulation study, missingness is induced artificially in real-world datasets, according to a mechanism. Experiments are conducted using 12 datasets from the UCI machine learning repository~\citep{asuncion2007uci} and their proposed method is compared with 8 imputation methods. The performance is measured using the RMSE calculated over imputed entries. Four missingness mechanisms are simulated, which are described below:
\begin{enumerate}
    \item \textit{MCAR}: By randomly masking data entries (\autoref{fig:m-graph mcar}).
    \item \textit{MAR}: A special version of Missing-At-Random mechanism where a number of variables are selected to be fully observed, and missingness in the remaining ones is induced as a function of them (based on a logistic model) (\autoref{fig:m-graph mar})
    \item \textit{MNAR}: Taking the MAR setting and randomly inducing missingness in fully-observed variables, now the missingness in remaining ones is dependant to partially observed data, hence the mechanism is Missing-Not-At-Random (\autoref{fig:m-graph mnar})
    \item \textit{self-censoring MNAR}: a logistic function of the each variable is used to induce missingness in itself (\autoref{fig:m-graph sc})
\end{enumerate}
For the logistic functions, the selected coefficients satisfy a `unity vector' restriction to avoid meaningless data synthesis. Analysis is done for different missingness ratios (10\%, 30\%, 50\%), different observed nodes for MAR and MNAR scenario, and different sample sizes. For all experiments, the number of iterations is $N_{\text{it}}=10$.

\subsubsection{PARCS simulation setting}

\autoref{tab:hyperimpute ext} describes the extension scenarios for the Hyperimpute simulation study. To focus more on showcasing PARCS utility, we run the experiments for only the compression dataset~\citep{yeh1998modeling}, covering MAR and MNAR settings with 50\% missingness ratio, and for two imputation methods, Hyperimpute and Missforest~\citep{stekhoven2012missforest}. We follow the paper's choice of iteration number, unless mentioned otherwise.

\begin{table}
  \centering
  \caption{Hyperimpute extension scenarios}
  \label{tab:hyperimpute ext}
  \begin{tabular}{cccl}
    \toprule
    Mechanism & Experiment no. & Extension & Description\\
    \midrule
    \multirow{4}{*}{MAR} & 1 & coefficient vector & Same mechanism, Wider range of coefficient vector \\
    & 2 & edge density & Same mechanism, fewer MAR edges \\
    & 3 & $R\rightarrow R$ edges & Different mechanism by adding new MAR edges \\
    & 4 & nonlinearity & Same mechanism, nonlinear structural equations \\
    \midrule
    MNAR & 5 & mechanism & Inducing other MNAR mechanisms \\
    \bottomrule
\end{tabular}
\end{table}
Implementation is done in three steps:
\begin{enumerate}
    \item \textit{$\mathcal{G}_Z$ subgraph}: We represent the dataset variables as graph nodes, which may only have out-going edges..
    
    \item \textit{$\mathcal{G}_R$ subgraph}: We specify the DAG and SCM models for the missingness indicator subgraph, partially and based on the intended missingness scenario.
    
    \item \textit{$\mathcal{G}_Z$-to-$\mathcal{G}_R$ edges}: We create missingness mechanism edges by randomizing the $Z \rightarrow R$ edges. We control the pattern of the connection by a particular `mask' per each mechanism.
\end{enumerate}

\begin{figure}
    \centering
    \begin{subfigure}[b]{0.45\textwidth}
        \centering
        \includegraphics[width=\textwidth]{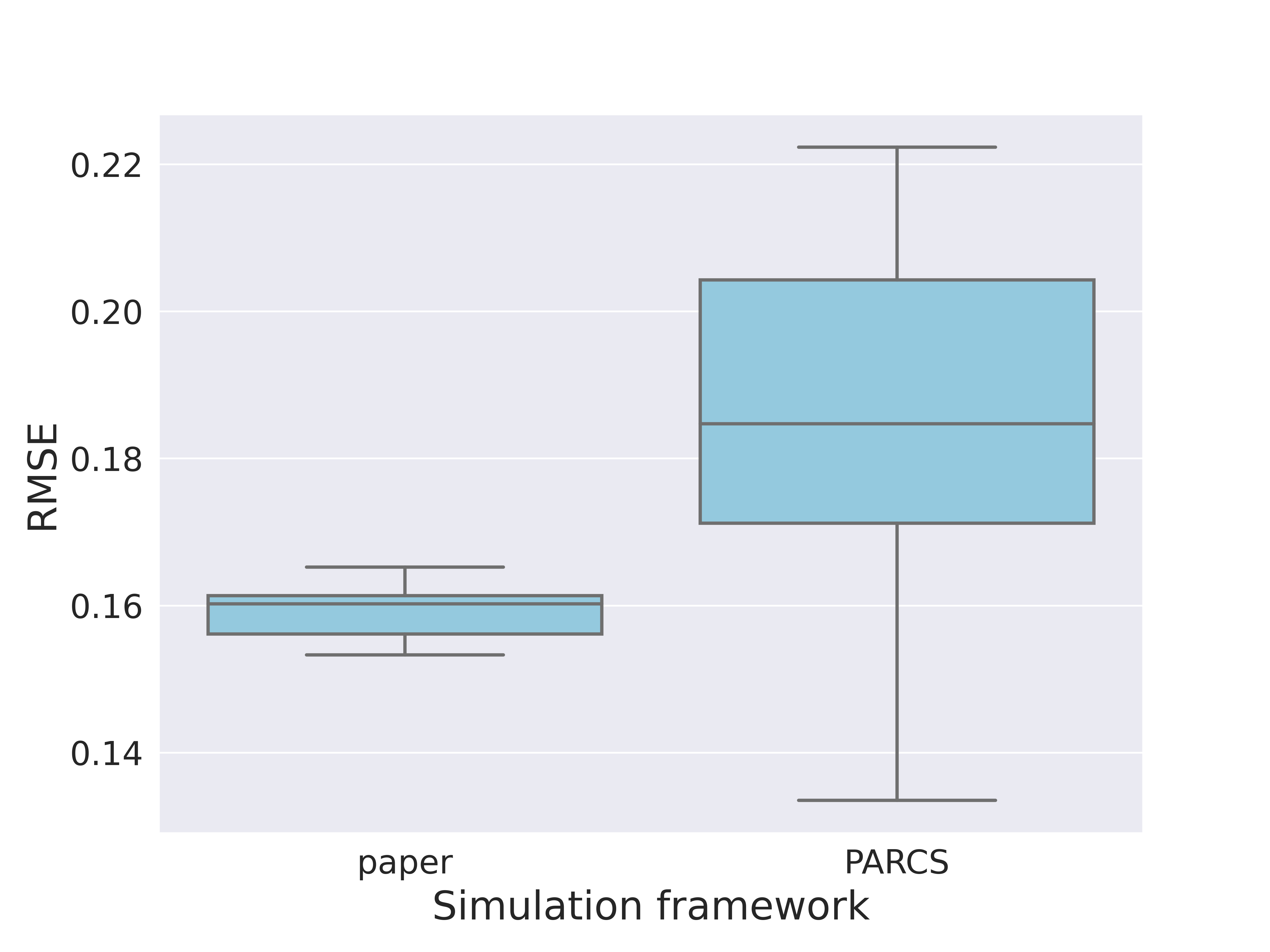}
        \caption{Reproducing the original MAR simulation with PARCS}
        \label{fig:miss results same mar}
    \end{subfigure}
    
    \begin{subfigure}[b]{0.48\textwidth}
        \centering
        \includegraphics[width=\textwidth]{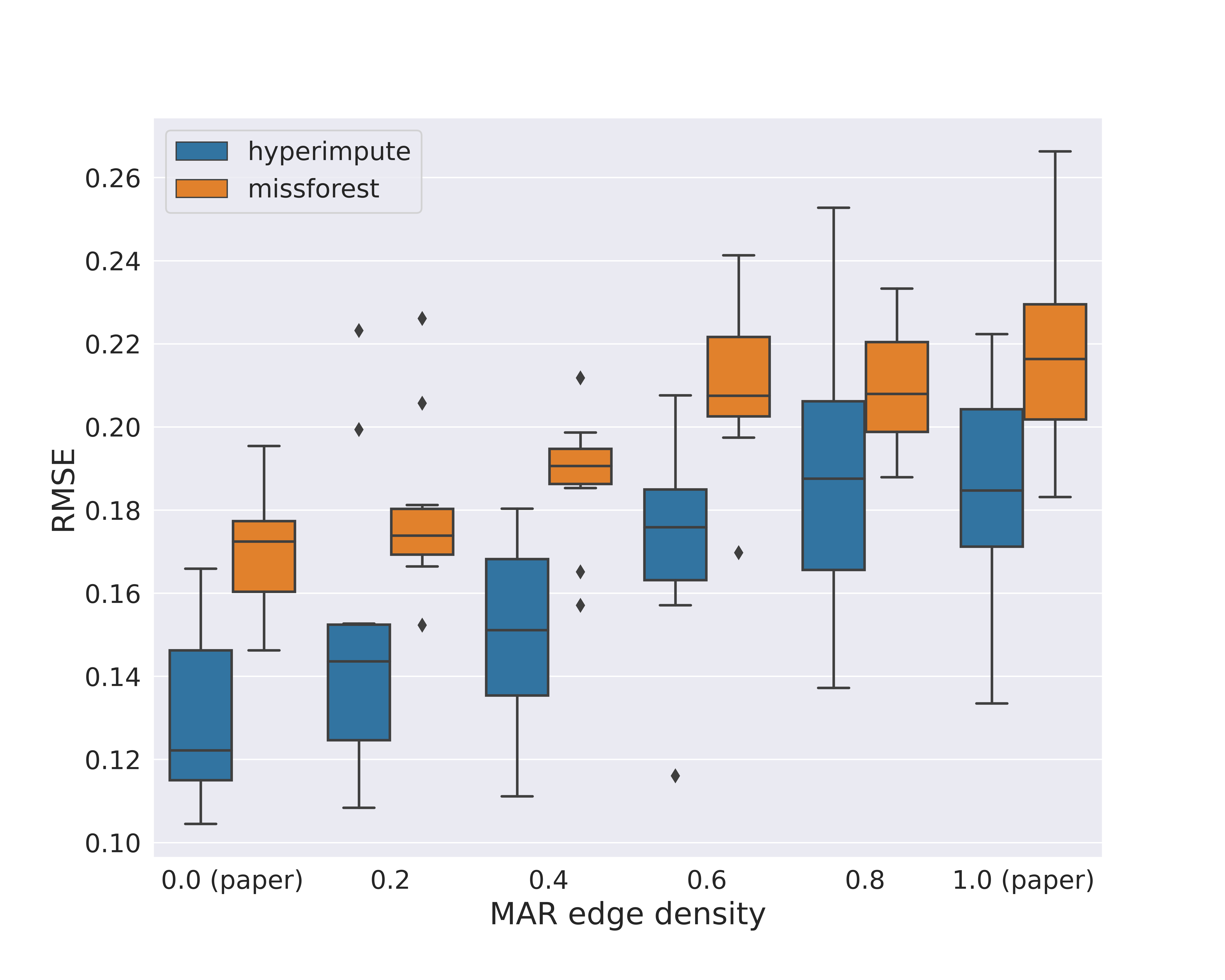}
        \caption{MCAR-to-MAR transition performance analysis}
        \label{fig:miss results same mcar-mar}
    \end{subfigure}
    \hfill
    \begin{subfigure}[b]{0.48\textwidth}
        \centering
        \includegraphics[width=\textwidth]{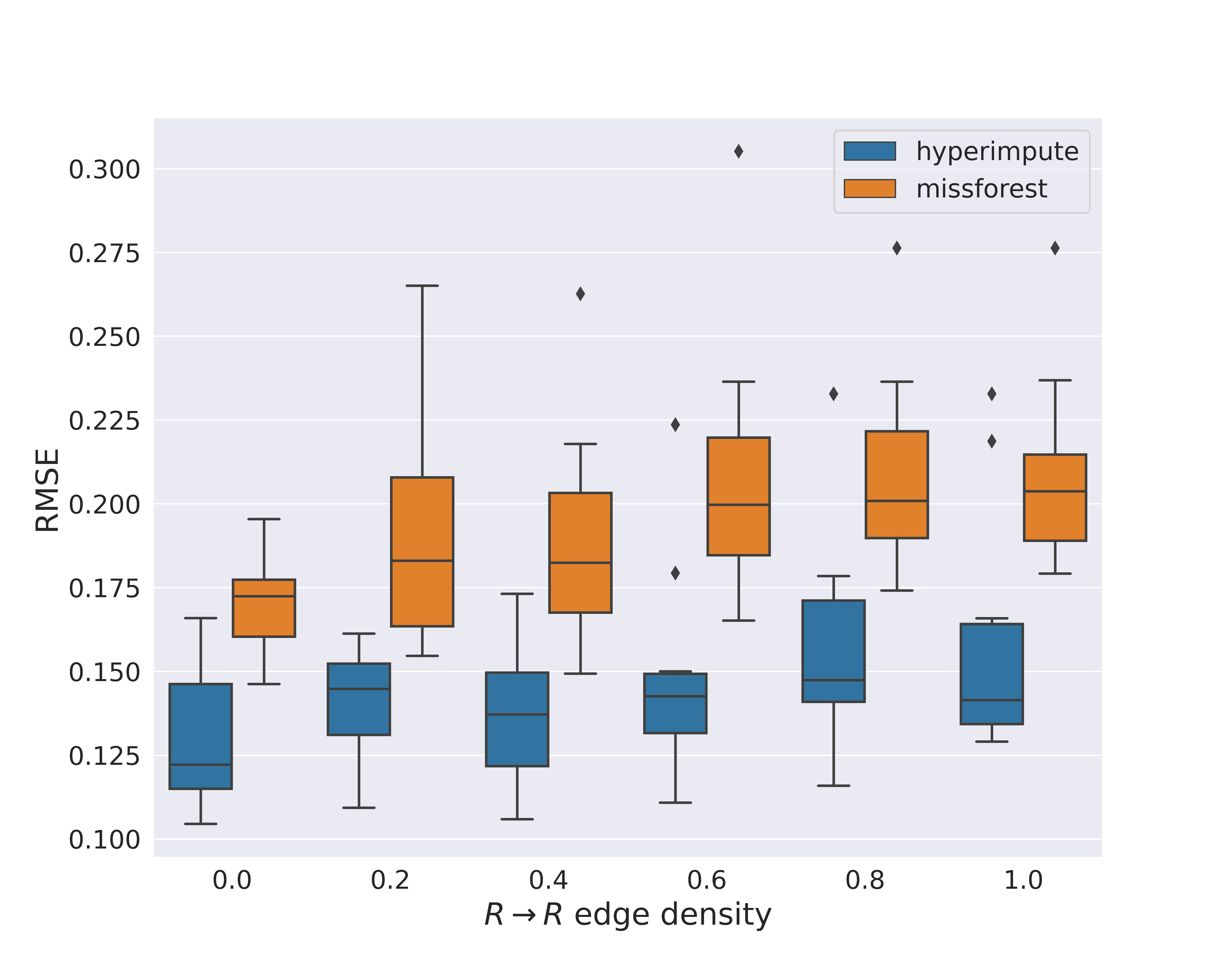}
        \caption{Inducing $R\rightarrow R$ edges in MAR}
        \label{fig:miss results same rr-mar}
    \end{subfigure}
    
    \begin{subfigure}[b]{0.48\textwidth}
        \centering
        \includegraphics[width=\textwidth]{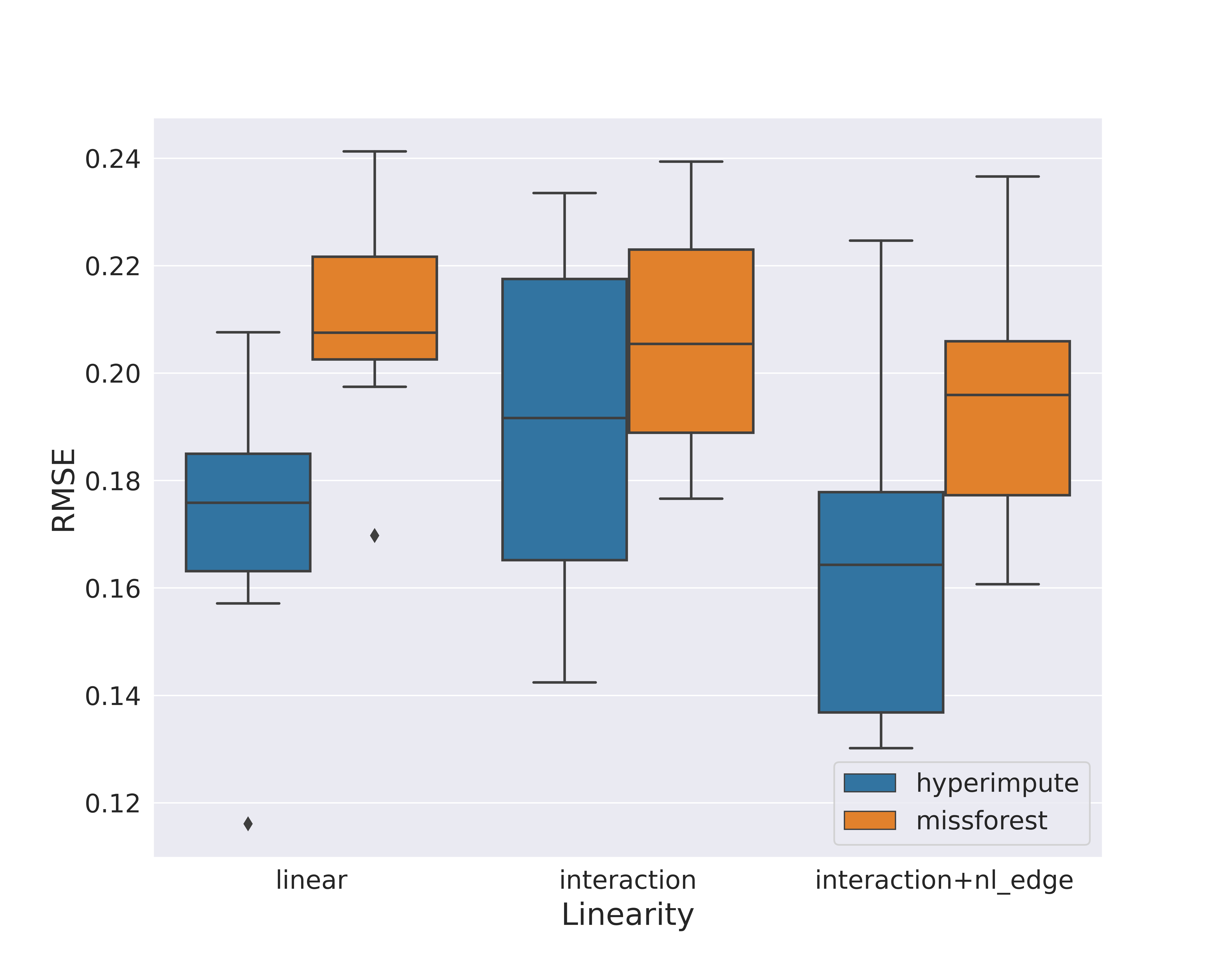}
        \caption{linear and nonlinear MAR scenarios}
        \label{fig:miss results same nonlin-mar}
    \end{subfigure}
    \hfill
    \begin{subfigure}[b]{0.48\textwidth}
        \centering
        \includegraphics[width=\textwidth]{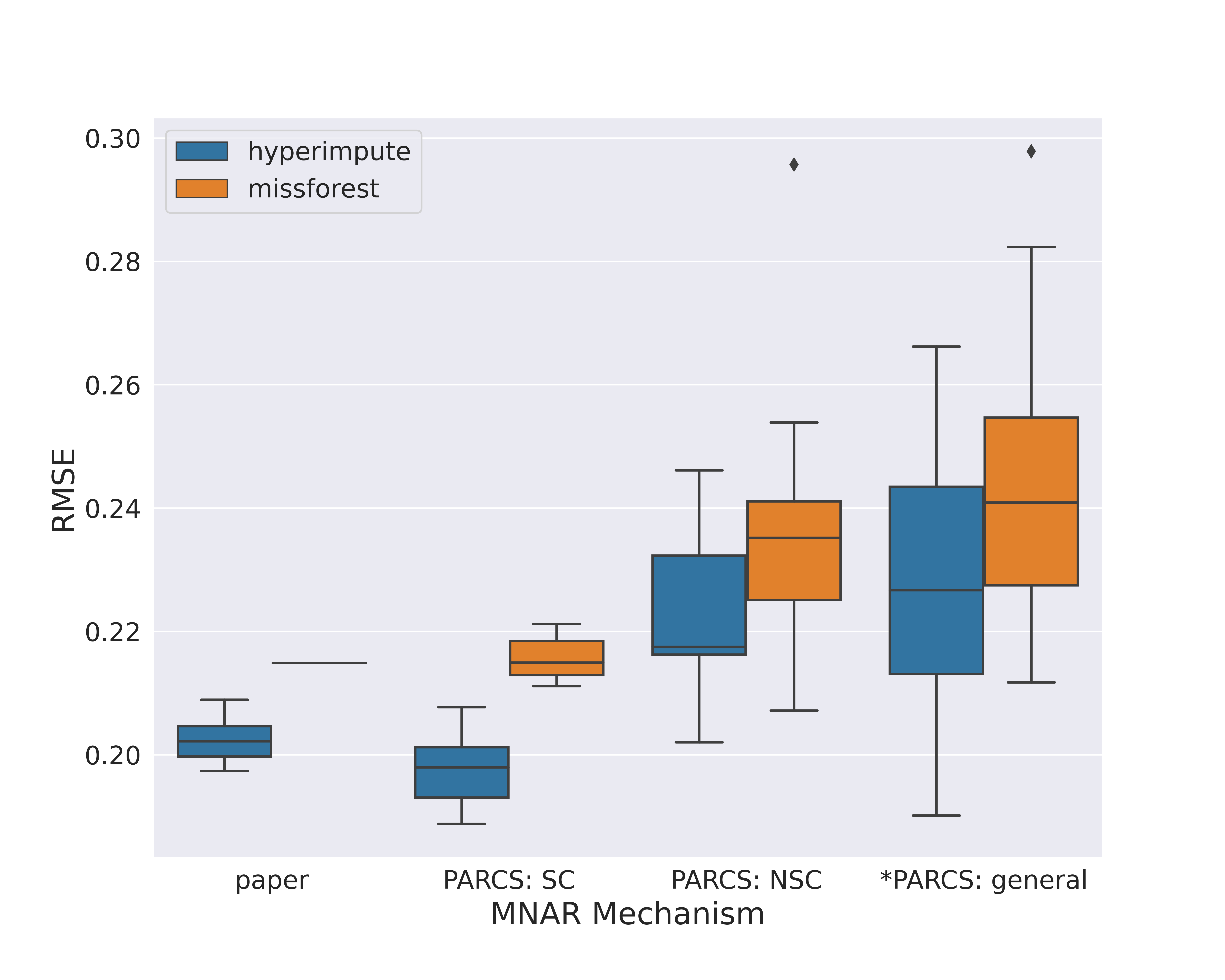}
        \caption{Inducing various MNAR scenarios}
        \label{fig:miss results same mnar}
    \end{subfigure}
    \caption{Simulation results for imputation methods, as done in the Hyperimpute paper \citep{jarrett2022hyperimpute}. Boxplots are obtained via 10 experiment iterations, except for the `*PARCS: general' column in Figure (e) with 30 iterations.}
    
    \label{fig:miss results}
\end{figure}

\paragraph{Experiment 1: Reproducing MAR results} Firstly, we produce the same MAR scenario in the paper (\autoref{fig:m-graph mar}). The only difference between the experiments is the selected ranges for coefficient vectors $W_p$ (weights of the success probability of the Bernoulli distribution for $R$ nodes): In the Hyperimpute paper, the coefficients are selected such that the $W$ becomes a unit vector. We randomize the coefficients in $[-5, -1] \cup [1, 5]$ (correction applied). \autoref{fig:miss results same mar} presents the RMSE boxplots for two simulation methods, where a considerable difference in both mean and standard deviation of the obtained RMSE is visible. This difference shows that the simulation in paper was incomplete due to free parameter limitations.

\paragraph{Experiment 2: MCAR-to-MAR transition} In the MAR m-graph in \autoref{fig:m-graph mar}, removing $Z \rightarrow R$ edges changes the mechanism to MCAR. Both MCAR and MAR scenarios are studied in the Hyperimpute paper. However, we can extend the analyses by visualizing the gradual increase in RMSE when the mechanism alters to MAR. This is achieved by varying the sparsity parameter of the randomization guideline from $0.0$ (MCAR) to $1.0$ (MAR). \autoref{fig:miss results same mcar-mar} presents the resulting boxplots for two imputation methods. Only first and last columns are reported by the paper (reproduced here).

\paragraph{Experiment 3: Extending to different MAR mechanism} \autoref{fig:miss results same rr-mar} presents the boxplot result for another MAR scenario which is not tested in the Hyperimpute paper. Here, the MAR mechanism is induced when $R$ nodes have edges among them too. This scenario is achieved by updating the $\mathcal{G}_R$ description with new edges. The results are generated for varying graph sparsity values. For this experiment, the $Z\rightarrow R$ density is set to $0.0$. Except the first column (corresponding to the paper's MCAR), the plot provides new performance measures in scenarios. Imputation methods show to be relatively robust to this extension of scenario.

\paragraph{Experiment 4: Non-linearity extension for MAR} Artificial missingness in the Hyperimpute paper is induced using a simple logistic function. from PARCS perspective, this means the coefficients for non-linear components in $\zeta$ are zero. In addition, only the identity edge function is used. We test the imputation methods in the presence of non-linear relations and interactions. \autoref{fig:miss results same nonlin-mar} presents the boxplot result of two imputation methods for linear and nonlinear relations. For this experiment, $R \rightarrow R$ and $Z \rightarrow R$ edge densities are selected $0.0$ and $0.6$ respectively. The first column (linear) presents the reproduced results, while last two columns demonstrate the methods' performance in untested scenarios. Similar to experiment 1, extension of simulation has lead to different RMSE scores, meaning that the reported performance in paper was limited by the simulation studies.

\paragraph{Experiment 5: Extending MNAR simulation} In the Hyperimpute paper, MNAR mechanisms are induced by either self-censoring or applying MCAR missingness to fully observed nodes of a MAR mechanism. Using PARCS, we simulate many known MNAR mechanisms by their corresponding m-graphs. These mechanisms are specified by the $Z\rightarrow R$ mask parameter. In this experiment we tested three scenarios: a general MNAR scenario (all-one mask) and two specific self-censoring and no-self-censoring scenarios. Note that SC and NSC scenarios are especial cases of the general scenario. \autoref{fig:miss results same mnar} presents the boxplot results of 3 scenarios alongside the result of the simulation done in the Hyperimpute paper. For general MNAR scenario, we ran the simulations for 30 iterations to achieve more reliable results, as the graph search space in this scenario is broader compared to other scenarios. The significant difference between the first boxplot (reproduced) and the other three (extended) points to the fact that the reported performance in the paper for MNAR missingness is limited by the incomplete simulation study.

\subsection{LiNGAM: a causal inference showcase}
LiNGAM is a well studied statistical causal model introduced by \citet{shimizu2006linear} alongside a causal inference and discovery method DirectLiNGAM. The model can shortly be written as $Z = BZ + \varepsilon$ with a lower diagonal matrix $B$ and independent, non-Gaussian error terms $\varepsilon$. Due to the linearity of the statistical model, DirectLiNGAM determines the causal order of the variables by recursively running linear regressions and removing the source nodes. The parameters of the estimated adjacency matrix $\hat{B}$ are then straightforward set as the estimated parameters of the corresponding regressions.

\subsubsection{Original Simulation Setting}
Using PARCS, we first implemented a class of distributions following the original paper the statistical model. In particular, we set the dimension of $X$ to $p = 5$, selected the adjacency matrix weights uniformly from the intervals $[-2, -0.5]$ and $[0.5, 2]$, forced the independent error terms to be log-exponentially distributed and permuted the entries of $X$ uniformly over $S_5$ to achieve a random causal ordering. As for the following two experiments, we then generated $500$ datasets with a dataset size of $1000$ samples each. As an objective to quantify the performance of DirectLiNGAM, we chose the Frobenius norm between the estimated and true adjacency matrix. The results can be seen in Figure \ref{fig:LiNGAM_frobenius} and there, smaller values on the y-axis indicate a better performance. As expected and shown in the green boxplot, DirectLiNGAM recovered the $500$ scenarios generated under the setting of the original paper very well.

\subsubsection{PARCS simulation setting}
DirectLiNGAM works good on data following the LiNGAM statistical model, which is however hardly ever satisfied for real data. Still, practitioners use and thus it is important to study its performance under model misspecification. With PARCS, we conduct two experiments altering the original setup resulting in datasets which are violating the linearity assumption.

\begin{figure}
    \centering
    \includegraphics[width = 0.8\textwidth]{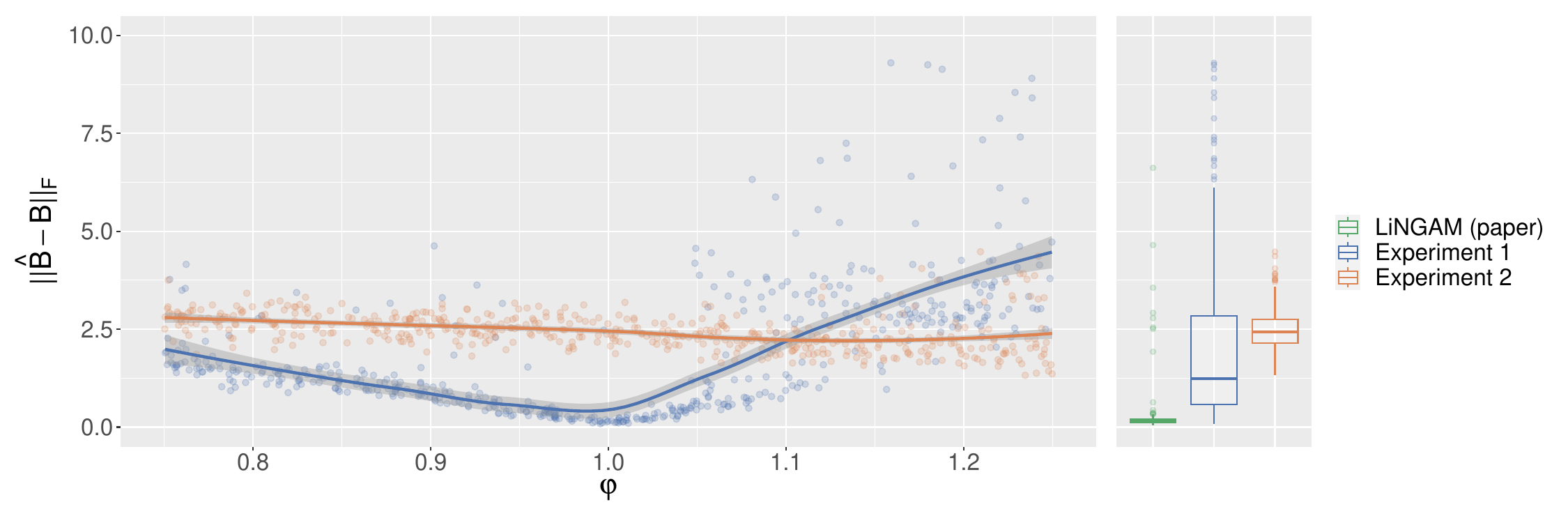}
    \caption{Frobenius norm of the difference between the estimated and true adjacency matrix $\hat{B} - B$ against the nonlinearity parameter $\phi$. Each boxplot in the right panel summarizes the performance over 500 datasets.}
    
    \label{fig:LiNGAM_frobenius}
\end{figure}

\paragraph{Experiment 1:} For the first alteration, we uniformly sample a constant $\phi \in [0.75, 1.25]$ and alter the structural equations from $Z_j = B_{j,:}(Z_1, \dots, Z_p)^\top + \varepsilon_j$ to \[ Z_j = B_{j,:}(g_\phi(Z_1), \dots, g_\phi(Z_p))^\top + \varepsilon_j \] for $j \in [p]$. The function $g_\varphi(z) = \text{sgn}(z) \cdot |z|^\phi$ introduces a nonlinear behaviour. In this setup, the value $|\phi - 1|$ quantifies the distance to the LiNGAM assumption and only datasets with $\phi = 1$ do follow the original statistical model. One implication by this data generating process is, that in all cases with $\phi < 1$, the variance of the variables will shrink with the causal order. The performance results of DirectLiNGAM is presented by the blue visualizations in Figure \ref{fig:LiNGAM_frobenius}. There we find a clear increase in recovery performance for $\phi$ near $1$. The further away $\phi$ is from $1$, the higher the model violation is, the worse is the recovery performance.

\paragraph{Experiment 2:} To counteract the unstable variance behaviour from the proceeding experiment, we add here a PARCS' edge correction as in \autoref{formula_correction} to all edges of the nonlinear data generating process.We find in formulae for all $j \in [p]$ the structural equations \[  Z_j = B_{j,:}(\frac{g_\phi(Z_1) - \mathrm{E}[g_\phi(Z_1)]}{\sqrt{\mathrm{Var}(g_\phi(Z_1))}}, \dots,\frac{g_\phi(Z_p) - \mathrm{E}[g_\phi(Z_p)]}{\sqrt{\mathrm{Var}(g_\phi(Z_p))}})^\top + \varepsilon_j. \] Although this setup stabilizes the variance, the relation between the variables is still nonlinear and thus, the data does not follow LiNGAM. This can in particular be also seen in the orange part of Figure \ref{fig:LiNGAM_frobenius} as we miss the drop for $\phi$ values near~$1$. Interestingly however, the general performance of DirectLiNGAM is better on the corrected datasets for $\phi > 1.1$ as the marginal distributions of e.g. $Z_5$ do not have a large variances and are instead compact.
Regarding the role of PARCS, it provided us a algorithmic-inexpensive option to test a causal discovery and inference method on a considerably extended distribution space. Especially for practitioners, it is important to also be able to trust a method when the underlying model is not correctly specified, and knowing how to analyse the observed distribution to assess the potential model performances.

\section{Conclusion} \label{sec:conclusion}

For causal inference and discovery methods, simulation studies provide reliable and comprehensive results only if they are designed according to the promised operational conditions of the methods. Many causal inference literature tend to over-restrict the design of their simulation study; this leads to unreliable and misleading simulation results.

PARCS simulation framework is designed to be effortlessly adjusted by method assumptions. It adopts the causal DAG and SCM models and provides a parameterization scheme to control the statistical and causal characteristics, and complexity of the data, from various aspects transparently and independently. As a result of this design, PARCS allows the users to partially fix a subset of parameters according to their assumptions, and sample the free parameters in iteration, to generate series of complying data generating processes.

In this paper, we introduced the simulation meta-algorithm of PARCS, and exemplified its effectiveness by reproducing and extending the simulation studies in two missing data analysis and causal discovery research papers. We showed that using PARCS, their empirical validation results would be more comprehensive.

We also compiled a list of desiderata for an effective causal simulation framework. We revisit the list and discuss how PARCS satisfies all the items:
\begin{itemize}
    \item \textit{Underlying causality}: PARCS models the underlying DGP with the widely-used concept of causal DAG and SCM models. The simulation studies are defined by specifying the DAG structure and SCM parameters.
    
    \item \textit{Data dimension and modality}: Arbitrary distributions can be assigned in order to simulate different data modalities. In addition, more complex SCMs can be simulated using intermediate nodes. On the data set level, sample size and number of features are adjustable, and directly controlled by the user.

    \item \textit{Controlled complexity}: PARCS controls the data complexity using the number of graph edges (sparsity), diverse output distributions, non-linear edge functions, and extended input vector dictionaries. Each of these sources can be adjusted separately. In all cases, partial randomization can be applied.

    \item \textit{Noise models}: The noise models are easily and independently adjusted by adopting different node distributions. Any sort of distribution can be simulated as long as the (parametric or non-parametric) iCDF of the distribution is defined.
\end{itemize}
By introducing the theoretical background of PARCS and publishing the Python package for public use, we propound the framework to the CI community for the future research in this field. PARCS adheres to the widely-used causal inference models, and facilitates transparent, reproducible, systematic, and comprehensive empirical investigations; thus we believe it is capable of playing an effective role as a benchmarking tool for CI methodologies.

\section{Acknowledgements}

This work was funded by the Bavarian Ministry for Economic Affairs, Regional Development and Energy as part of a project to support the thematic development of the Institute for Cognitive Systems.

We thank Octavia Ciora and Elisabeth Pachl for proofreading and providing instructive feedback.

\bibliographystyle{ACM-Reference-Format}
\bibliography{references}

\end{document}